\documentclass  [twocolumn, amsmath,amssymb,aps,showpacs, superscriptaddress]{revtex4-1}

\usepackage{graphicx}
\usepackage{dcolumn}
\usepackage{bm}
\usepackage{braket,amsmath,amssymb,bm}
\usepackage{color}
\usepackage{siunitx}
\usepackage{float}
\usepackage{hyperref}
\usepackage{lineno}
\usepackage{multirow}
\usepackage{tabularx}
\usepackage{enumitem}

\begin{document}

\title{Real-time estimation of the optically detected magnetic resonance shift in diamond quantum thermometry}

\author{Masazumi Fujiwara}
 \email{masazumi@osaka-cu.ac.jp}
 \affiliation{Department of Chemistry, Osaka City University, Sumiyoshi-ku, Osaka, 558-8585, Japan}
 
\author{Alexander Dohms}
 \affiliation{Institut f{\" u}r Physik und IRIS Adlershof, Humboldt Universit{\" a}t zu Berlin, Newtonstrasse 15, 12489 Berlin, Germany}

\author{Ken Suto}
 \affiliation{Department of Chemistry, Osaka City University, Sumiyoshi-ku, Osaka, 558-8585, Japan}
 
\author{Yushi Nishimura}
 \affiliation{Department of Chemistry, Osaka City University, Sumiyoshi-ku, Osaka, 558-8585, Japan}

\author{Keisuke Oshimi}
 \affiliation{Department of Chemistry, Osaka City University, Sumiyoshi-ku, Osaka, 558-8585, Japan}

\author{Yoshio Teki}
 \affiliation{Department of Chemistry, Osaka City University, Sumiyoshi-ku, Osaka, 558-8585, Japan}
 
\author{Kai Cai}
 \affiliation{Department of Electrical and Information Engineering, 
 Osaka City University, Sumiyoshi-ku, Osaka, 558-8585, Japan}

\author{Oliver Benson}
 \affiliation{Institut f{\" u}r Physik und IRIS Adlershof, Humboldt Universit{\" a}t zu Berlin, Newtonstrasse 15, 12489 Berlin, Germany}
 
\author{Yutaka Shikano}
 \email{yutaka.shikano@keio.jp}
 \affiliation{Quantum Computing Center, Keio University, 3-14-1 Hiyoshi, Kohoku, Yokohama, 223-8522, Japan}
 \affiliation{Institute for Quantum Studies, Chapman University, 1 University Dr., Orange, CA 92866, USA}

\begin{abstract}
We investigate the real-time estimation protocols for the frequency shift of optically detected magnetic resonance (ODMR) of nitrogen-vacancy (NV) centers in nanodiamonds (NDs). 
Efficiently integrating multipoint ODMR measurements and ND particle tracking into fluorescence microscopy has recently demonstrated stable monitoring of the temperature inside living animals. 
We analyze the multipoint ODMR measurement techniques (3-, 4-, and 6-point methods) in detail and quantify the amount of measurement artifact owing to several systematic errors derived from instrumental errors of experimental hardware and ODMR spectral shape.
We propose a practical approach to minimize the effect of these factors, which allows for measuring accurate temperatures of single NDs during dynamic thermal events. 
We also discuss integration of noise filters, data estimation protocols, and possible artifacts for further developments in real-time temperature estimation.
The present study provides technical details of quantum diamond thermometry and discusses factors that may affect the temperature estimation in biological applications.
\end{abstract}
\maketitle

\section{Introduction}
Quantum nanoscale sensing based on optically detected magnetic resonance (ODMR) of nitrogen-vacancy (NV) centers in nanodiamonds (NDs) enables highly sensitive nanoscale probing of magnetic fields, electric fields, and temperature~\cite{maze2008nanoscale,schirhagl2014nitrogen,RevModPhys.89.035002,petrini2020quantum}. 
Reading the frequency shift of ODMR using continuous-wave (CW) or pulsed measurements is the fundamental step. 
Recent studies have focused more on developing this sensing technique for practical applications~\cite{doi:10.1002/2017GC006946,rendler2018fluorescent,levine2019principles,Lesik1359,CHOE20181066,doi:10.1021/acs.nanolett.8b00895} and have accordingly tried to efficiently deploy the sensors into real-time measurement systems~\cite{PhysRevLett.106.030802,bonato2016optimized,rendler2017,doi:10.1063/1.5010282,PhysRevApplied.10.034044,Ambal_2019}. 

For biological applications, the use for thermometry is of particular significance because temperature is a fundamental parameter of biological activity, such as circadian rhythms~\cite{akin2011homeostatic}, energy metabolism~\cite{10.3389/fphys.2017.00520}, and developmental processes~\cite{BEGASSE2015647,chong2018temporal}.
The biological application of NV thermometry was first demonstrated in cultured cells~\cite{kucsko2013nanometre,simpson2017non,yukawa2020quantum,sekiguchi2018fluorescent}, and recently in \textit{in-vivo} model animals, such as nematode worms~\cite{choi2019,fujiwara2019realtime}.
A technological key to the recent \textit{in-vivo} demonstrations is the method to efficiently determine the temperature of NDs under their dynamic motion inside biological structures; one needs to complete the temperature measurement for a few seconds because the incorporated NDs and whole systems are moving, which significantly challenges the current quantum-sensing protocols.
Of various ODMR measurement protocols related to thermometry~\cite{kucsko2013nanometre,neumann2013high,Toyli8417,tzeng2015time,clevenson2015broadband,PhysRevB.91.155404,PhysRevX.8.011042,liu2019coherent,Liu2019arXiv}, multipoint methods have been proposed ~\cite{PhysRevLett.106.030802,kucsko2013nanometre,tzeng2015time,PhysRevX.8.011042} to accelerate the ODMR measurement process, wherein fluorescence intensities at three or four frequency points are acquired to estimate the ODMR shift. 
These methods significantly reduce the measurement time and thus enable more signal sampling, which improves the precision, on comparing with the frequency-shift determination using the whole-spectral-shape measurements. 
These methods, however, are inevitably susceptible to experimental errors, such as change of ODMR spectral shape, temperature-dependent NV fluorescence intensity, and hardware's instrumental errors, 
because they \textit{estimate} the frequency shift based on the limited available information at the chosen frequency points.

These two types of multipoint ODMR methods, i.e., 3- and 4-point methods, have demonstrated their effectiveness ~\cite{kucsko2013nanometre,tzeng2015time}.
Intuitively, the number of frequency points is likely to affect the results; the estimation based on a higher number of frequency points provides more information and shows a robust behavior under the complicated response  of ODMR spectral shape that cannot be simplified using the simple Lorentzian function.
However, increasing the number of frequency points reduces the number of measurements, which may lead to deterioration in precision. 
Understanding error and noise sources associated with these multipoint ODMR measurements is crucial to use these techniques further.

In this report, we study the multipoint ODMR measurements in the context of real-time thermometry for biological applications. 
Focusing on the 4-point method, we first show that there is a marginal difference in the photo-responsivity between the photon counts of the selected four frequencies, which causes  significant artifacts during temperature estimation. 
This difference in photo-responsivity is also observed for the 3-point method, which can be attributed to the variations of ODMR spectral shape and temperature-dependent NV fluorescence intensities.
In addition, the hardware's instrumental errors can affect the photo-responsivity, which becomes significant when using the 6-point method (an extended version of the 4-point method).
Second, we propose a practical way to cancel this effect, and compare the performances of each method in terms of precision and associated noises.
Third, we demonstrate the monitoring of temperature dynamics of single NDs while having dynamic variation of fluorescence intensity by the thermal drift of the setup and temperature dependence of NV fluorescence.
Finally, we summarize the above-detailed investigation and discuss possible artifacts that may be produced while using multipoint ODMR measurement techniques on biological samples.

\begin{figure*}[th!]
 \centering
 \includegraphics[scale=0.9]{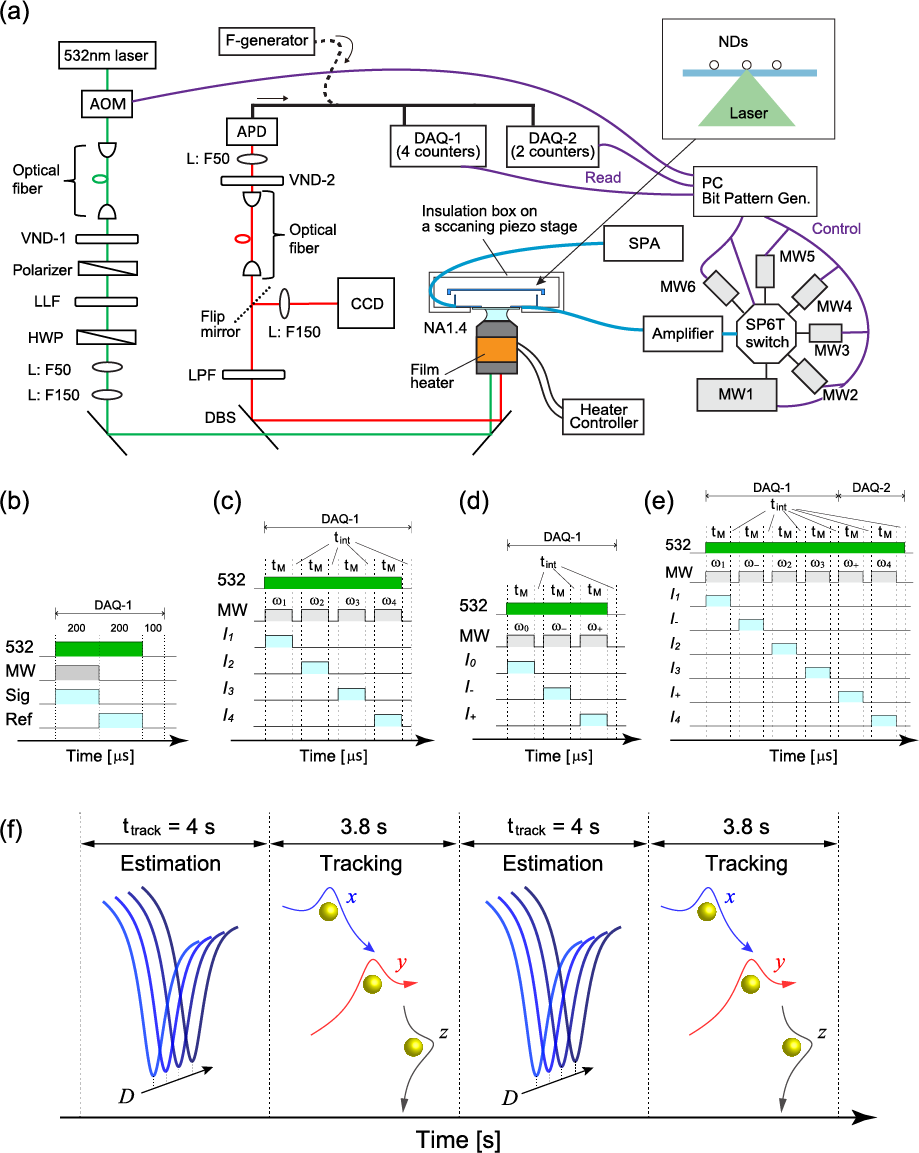}
 \caption{(a) A schematic drawing of the experimental setup for the optical layout and microwave circuit. AOM: acousto-optic modulator. VND: variable neutral density filter. LLF: laser-line filter. HWP: half-wave plate. L: lens. DBS: dichroic beam splitter. LPF: long-pass filter. CCD: charge-coupled device camera. APD: avalanche photodiode. SPA: spectrum analyzer. MW: microwave source. DAQ: data-acquisition board. F-generator: frequency generator. Inset drawing: NDs spin-coated on a coverslip. (b) Pulse control sequences for CW-ODMR measurements and (c, d, e) 4-, 3-, and 6-point measurements. 532: green laser. MW: microwave. Sig: signal for $I_{\rm PL}^{\rm ON}$. Ref: reference for $I_{\rm PL}^{\rm OFF}$. $\omega_1$ to $\omega_4$ are the four frequencies used for the 4-point measurements. $\omega_-$, $\omega_+$ are the additional frequencies used for 6-point measurements. For the 3-point measurement, $\omega_-$, $\omega_+$ and off-resonant $\omega_0$ (2.65 GHz) were used. $I_1$ to $I_4$ and $I_-$, $I_+$ are the corresponding photon counts. $t_{\rm M}$: measurement time. $t_{\rm int}$: interval time. (f) A sequence of the estimation of the zero-field splitting ($D$) and three-dimensional particle tracking in the $xyz$ directions. $t_{\rm track}$: tracking period.
 }
 \label{fig1}
\end{figure*}

\section{Experiments} 
\subsection{CW and multipoint ODMR measurements}

\begin{figure}[t!]
 \centering
 \includegraphics[scale=1.0]{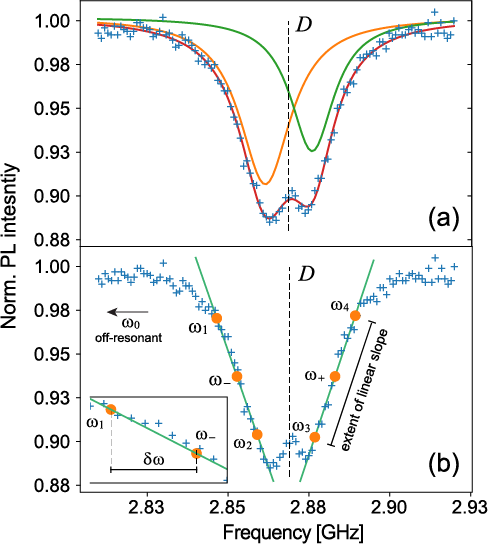} 
 \caption{(a) ODMR spectrum fitted using the sum of two Lorentzian functions centered at the zero-field splitting frequency $D$ and (b) that with two linear slopes fitted to the spectrum. The selected frequencies $\omega_1$ to $\omega_4$ and $\omega _-$, $\omega _+$ are represented as orange circles from left to right, and are separated on each of the slopes by $\delta \omega$, as shown in the inset. $\omega_0$ is used for the 3-point method and set to 2.65 GHz sufficiently far from $D$. }
 \label{fig2}
\end{figure}

Figure~\ref{fig1} shows a schematic diagram of the experimental setup and control sequences for the CW and multipoint ODMR measurements.
A CW 532-nm laser was used to excite the ND-NV centers, with an intensity of $\sim$ 2 kW$\cdot {\rm cm ^{-2}}$. 
An oil-immersion objective with a numerical aperture of 1.4 was used. 
The NV fluorescence was filtered using a dichroic beam splitter (Semrock, FF560-FDi01) and long-pass filters (Semrock, BLP01-635R-25) to remove the residual scattered signal of the green laser.
The fluorescence emission was coupled to an optical fiber that acted as a pinhole (Thorlabs, 1550HP).
The fiber-coupled fluorescence was detected by an avalanche photodiode (Excelitas, SPCM AQRH-14). 
The output of an APD was fed to two data-acquisition (DAQ) boards equipped with four pulse counters (DAQ-1: National Instruments, USB-6343 BNC) and two counters (DAQ-2: National Instruments, USB-6229 BNC). 
We used NDs containing $\sim$ 500 NV per particle (Ad\'{a}mas Nanotechnologies, NDNV100nmHi10ml), which were then spin coated on coverslips. 
The sample coverslips were set in an incubation chamber that was mounted on a piezo stage.
Raster scanning and particle tracking of NDs were performed by controlling the voltage applied to a piezo stage (Piezosystemjena, TRITOR-100SG).
Note that an external magnetic field was not used in this study.

To implement both CW and multipoint ODMR measurements, a stand-alone microwave source (Rohde \& Schwarz, SMB100A) and five USB-powered microwave sources (Texio, USG-LF44) were connected to an SP6T switch with a switching time of 250 ns (General Microwave, F9160).
The output microwave signal was amplified (Mini-circuit, ZHL-16W-43+) and fed to a microwave linear antenna placed on a coverslip (25-\si{\um}-thin copper wire) that was sealed with a home-built plastic box with a hole in the center.
The typical microwave excitation power was estimated to be 10--50 mW (10--17 dBm) at the linear antenna by considering the source output, amplifier gain, and the experimentally determined $S_{21}$ of the antenna system, which provides microwave magnetic field of more than 2--5 gauss in 20 $\si{\um}$ from the antenna.
To acquire CW-ODMR measurements, the APD detection was gated for microwave irradiation ON and OFF using the SP6T switch and a bit pattern generator (Spincore, PBESR-PRO-300).
Specifically, the bit-pattern generator fed TTL pulses to the SP6T switch for gating the output from MW1 (200 $\si{\us}$ for ON and OFF), which is followed by 100-$\si{\us}$ TTL pulse to the acousto-optic modulator for switching off the laser, as described in Fig.~\ref{fig1}(b).
We prepared two counters in DAQ-1 (Sig and Ref in Fig.~\ref{fig1}(b)) and fed the APD output to one of its digital input ports to be assigned to these two counters.
These two counters were operated in ”gated edge counting mode”.
This resulted in $I_{\rm PL}^{\rm ON}$ and $I_{\rm PL}^{\rm OFF}$ with a repetition rate of 2 kHz.
In the multipoint-ODMR measurements, the APD detection was gated for the respective microwave frequencies (4-points: $\omega_1$ to $\omega_4$. 3-points:  $\omega_-$, $\omega_+$, and off-resonant $\omega_0$. 6-points: $\omega_1$ to $\omega_4$, $\omega_-$, $\omega_+$). 
The gate width was $t_{\rm M} = 100 \ \si{\us}$, common to all the four, three, or six gates, each followed by an interval of $t_{\rm int} = 5 \ \si{\us}$. 
For the 4- or 3-point method, we used four or three counters in DAQ-1, each synchronized to the respective microwave sources (4-point: MW1 to MW4. 3-point: MW1 to MW3).
For the 6-point method, the four counters of DAQ-1 and two counters of DAQ-2 were each synchronized to the respective microwave sources (MW1 to MW6). 
The total photon count $I_{\rm tot}$ was obtained using the following equation:
\begin{equation}
    I_{\rm tot} = \frac{t_{\rm M}+t_{\rm int}}{t_{\rm M}} \times \sum_{k}^{4, 3, 6} I_k, 
\end{equation}
where 1.05 was used as the correction factor to account for the time interval of 5 $\si{\us}$, during which no photons were counted.

During the temperature measurements, a confocal microscope system was used to track the target NDs. 
The piezo stage was scanned in the $xyz$ directions while measuring the ND fluorescence intensity. 
The obtained cross-sections of the point spread function along the $xyz$ axes were fitted using Gaussian functions to determine the $xyz$ positions for re-positioning. 
The piezo stage was moved smoothly to the re-positioning point in five steps of $\sim$ 20 nm every 2 ms.
The re-positioning took 3.8 s and was performed with a tracking period of $t_{\rm track} = $ 4 s.
It should be noted that data acquisition for the temperature estimation was performed for 1.0 s (occasionally, for 0.5 s, depending on the application) for every four seconds.

It should be noted that there may be other combinations of experimental devices for using the multipoint methods, which may affect the interpretation of a part of the following results, particularly regarding the artifact caused due to experimental hardware.
For example, one could use arbitrary wave-form generator to make multipoint sequences instead of using both microwave switches and many microwave frequency sources. 
Devising the gate-feeding to the DAQ board may reduce the number of counters used in the DAQ boards by sharing the same counters for the multiple measurement windows. 
Field-programmable gate array (FPGA) may provide us with wider choices for the implementation of multipoint ODMR. 
Using such an implementation could eliminate some of the hardware-derived measurement artifacts.
However, the following section reports a measurement artifact that is most likely intrinsic to NV ODMR, that is, it cannot be removed by using the above-mentioned hardware implementations. 

\subsection{ODMR spectral analysis to define the frequency points for the multipoint ODMR measurements}
To determine the microwave frequencies for the multipoint ODMR measurements, the CW-ODMR spectra were first recorded for target NDs.
Based on the obtained CW-ODMR spectra, the intensities $I_1$ to $I_4$ and $I_-$, $I_+$ were selected in the following manner: 
(1) The obtained CW-ODMR spectra were fitted to the sum of two Lorentzian functions to indicate the ODMR dip, and the zero-field splitting $D$, as shown in Fig.~\ref{fig2}. 
(2) The two linear slopes of the ODMR dip were determined via linear fits and six frequency points that included three on each slope, which were uniformly distributed, i.e., equidistant with $\delta \omega$ over the extent of the slopes. 
In this case, $\omega_-$ and $\omega_+$ were centered on the zero-field splitting ($D$) such that $I(\omega_-)=I(\omega_+)$ (see Appendix for more details). 

Combinations of the fluorescence intensity values at these frequency points provide the ODMR shift estimation ($\Delta \Omega _{\rm NV}$) for the 4-, 3-, and 6-point methods. 
For the 4-point method~\cite{kucsko2013nanometre}, we have 
\begin{equation}
\begin{split}
&\Delta \Omega _{\rm 4pnt}^1 = \delta \omega \frac{(I_1+I_2) -(I_3+I_4)}{(I_1-I_2) - (I_3-I_4)}, 
\\
&\Delta T_{ \rm NV}^{\rm 4pnt} = \alpha^{-1} \Delta \Omega _{\rm 4pnt}^1, 
\end{split}
\label{eq:4pnt}  
\end{equation}
where $\alpha = dD/dT$ is the temperature dependence of the zero-field splitting that mainly originates from thermal lattice expansion~\cite{PhysRevLett.104.070801,doi:10.1063/1.3652910,PhysRevB.90.041201,PhysRevApplied.10.034009,fukami2019all-optical}.

For the 3-point method~\cite{tzeng2015time}, we take $\omega_-$ and $\omega_+$ for the two frequencies on the Lorentzian dip and set $\omega_0 = 2.65$ GHz for the off-resonance frequency to take the baseline fluorescence intensity. 
The temperature estimate can then be written as follows~\cite{tzeng2015time}:  
\begin{equation}
\begin{split}
\Delta \Omega _{\rm 3pnt} &= - \Gamma  \frac{1+\rho^2}{2\rho}\frac{I_+ - I_-}{2I_0 - I_+ - I_-},
\\
\Delta T_{\rm NV}^{\rm 3pnt} &= \alpha^{-1} \Delta \Omega _{\rm 3pnt}, 
\end{split}
\label{eq:3pnt} 
\end{equation}
where $\Gamma$ is the full width at half maximum (FWHM) of single Lorentzian and $\rho = |\omega_+ - \omega_-| / \Gamma$.
Note that the reason of choosing 2.65 GHz as an off-resonant microwave frequency is because it is sufficiently far from the resonant 2.87 GHz but is not so far as to change the high-frequency noise to the piezo stage or other electronic devices.

For the present 6-point method, we simply take the mean of the three types of estimates of the 4-point method as follows: 
\begin{equation}
\begin{split}
&\Delta \Omega _{\rm 4pnt}^1 = \delta \omega \frac{(I_1+I_2) -(I_3+I_4)}{(I_1-I_2) - (I_3-I_4)}, 
\\
&\Delta \Omega _{\rm 4pnt}^2 = \frac{\delta \omega}{2} \frac{(I_1+I_-) -(I_+ + I_4)}{(I_1-I_-) - (I_+ -I_4)}, 
\\ 
&\Delta \Omega _{\rm 4pnt}^3 = \frac{\delta \omega}{2} \frac{(I_- + I_2) - (I_3 + I_+)}{(I_- - I_2) - (I_3 - I_+)}. 
\end{split}
\label{eq:6pnt-1} 
\end{equation}
Then, we take their mean as
\begin{equation}
\begin{split}
\Delta \Omega _{\rm 6pnt} &= \frac{\Delta \Omega _{\rm 4pnt}^1 + \Delta \Omega _{\rm 4pnt}^2 + \Delta \Omega _{\rm 4pnt}^3}{3}, 
\\
\Delta T_{\rm NV}^{\rm 6pnt} &= \alpha^{-1} \Delta \Omega _{\rm 6pnt}.
\end{split}
\label{eq:6pnt-2} 
\end{equation}
From these equations, the measurement noise can be calculated. 

It should be noted that a comparison of the precision of these multipoint ODMR methods needs experimental validation, which is difficult theoretically.
Single sequences of the 4-, 3-, and 6-point measurements take $t_{\rm 4pnt} = 420 \ \si{\us}$, $t_{\rm 3pnt} = 315 \  \si{\us}$, and $t_{\rm 6pnt} = 630 \ \si{\us}$, respectively.
Supposing the photon count rate to be $R$, photon shot noise of the single measurement of $i$-point method ($\sigma$) can be written as $\sqrt{R \times t_{i \rm{pnt}}}$.
The measurement noise with 1-s sampling time is determined by the number of measurements per second ($t_{i \rm{pnt}}^{-1}$), which provides the measurement noise 
\begin{equation}
    \sigma_{i\rm{pnt}} = \sqrt{R} \times t_{i\rm{pnt}}.
    \label{eq:ipnt-noise} 
\end{equation}
Therefore, the 3-point method can provide the smallest precision in a simplistic picture because it can integrate as many measurements (estimation) as possible for a certain time.
However, there are other factors affecting the precision of the thermometry, such as spectral shape of the ODMR.
The assumption of a single Lorentzian shape is not straightforward because ODMR dip exhibits splitting due to the mixed contribution from the interference between the $\rm{^3A}$ spin states and from the inhomogeneous decoherence sources~\cite{matsuzaki2016optically,fujiwara2016manipulation,simpson2017non}.
Such spectral shape dependency will be discussed in Sec.~\ref{sec3A} in detail.
The 4- and 6-point methods do not explicitly use such an assumption of the spectral shape; however, the selection of frequency points affects the measurement. 
For example, frequency points on one side ($\omega_1$ and $\omega_2$ or $\omega_3$ and $\omega_4$) should be apart from each other to obtain smaller noise ($\sim \sigma_{\rm 4pnt}/\delta \omega$), but they need to be sufficiently far from the curved region of the ODMR spectrum, where the linear fit is no longer valid.

\subsection{Experimental determination of the precision and accuracy}
The precision ($\sigma_{\rm exp}$) of the thermometry was experimentally determined by taking the standard errors of 20 sampling points of ${\Delta T_{\rm NV}}$ that were recorded over 38 s, including tracking time. 
The sensitivity ($\eta_{T}$) could be calculated as $\eta_{T} = \sigma_{\rm exp} \times \sqrt{2 \delta t_{\rm intgr}}$ because this duration consists of 19.4-s integration time ($\delta t_{\rm intgr}$) and 18.6-s re-positioning time.
The accuracy was determined by adding an offset ($T_0$) to ${\Delta T_{\rm NV}}$ to match ${T_{\rm S}}$ and taking the root-mean-square (RMS) of ${T_{\rm S}-(T_0 + \Delta T_{\rm NV}}$), where $T_{\rm S}$ is the sample temperature calibrated separately (see below Sec.~\ref{subsec:external-T}).
The upper bound of the RMS in the steady state was considered to be the accuracy.
It should be noted that the fluctuation of the environmental temperature $T_{\rm air}$ (namely, the air-conditioned laboratory room temperature) caused a deviation of $T_0 + \Delta T_{\rm NV}$ from $T_{\rm S}$, which overestimated the accuracy value.

\begin{figure*}[t!]
 \centering
 \includegraphics[scale=1.1]{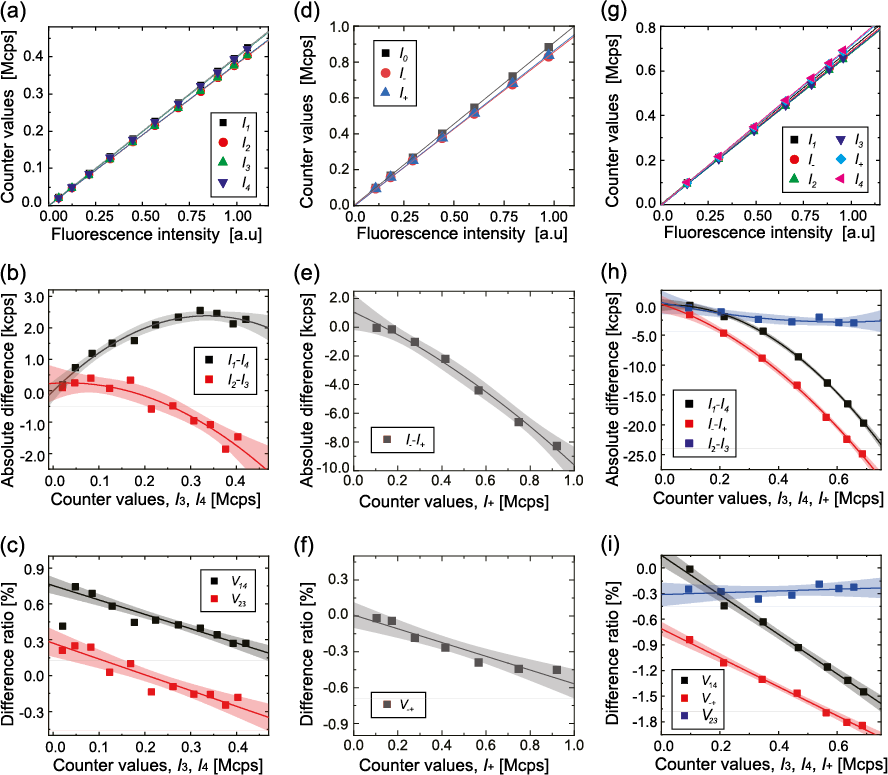}
 \caption{Variation of the photo-responsivity of the counters. (a, d, g) Photon counts for each counter as a function of ND fluorescence with their linear fits for the 4-, 3-, and 6-point measurements, respectively.
 (b, e, h) Difference in the counter values between the sets of two counters, namely,  $I_1 - I_4$, $I_2 - I_3$, as functions of $I_3$ and $I_4$ for the 4-point, $I_- - I_+$ as a function of $I_-$ for the 3-point, and $I_1 - I_4$, $I_2 - I_3$, $I_- - I_+$ as functions of $I_3$, $I_4$, and $I_-$ for the 6-point measurements, respectively. 
 The solid lines and shaded area represent second-order polynomial fits to the data and their 95-\% confidence intervals, respectively.
 (c, f, and i) The corresponding difference ratio of $V_{ij}$.
 The solid lines and shaded area represent linear fits to the data and their 95-\% confidence intervals, respectively.
 }
 \label{fig3}
\end{figure*}

\subsection{External temperature change}
\label{subsec:external-T}
The sample temperature ($T_{\rm S}$) was varied via direct heat conduction from the oil-immersion microscope objective, whose temperature ($T_{\rm obj}$) was controlled by using a PID-feedback controller of the foil heater which wrapped the objective (Thorlabs, HT10K \& TC200, temperature precision: $\pm$ 0.1 K).
The immersion oil was Olympus Type-F.
$T_{\rm S}$ was calibrated in the following manner: (1) a tiny flat Pt100 resistance temperature probe (Netsushin, NFR-CF2-0505-30-100S-1-2000PFA-A-4, $5 \times 5 \times 0.2 \ {\rm mm}^{3}$) was tightly attached to the sample coverslip by a thin layer of silicone vacuum grease between the probe and the coverslip. (2) The probe was completely covered with aluminum tape whose edges were glued to the base coverslip. (3) In this thermal configuration, $T_{\rm obj}$ was varied while monitoring $T_{\rm S}$. 
We obtained the following relation: $T_{\rm S} = 1.847 + 0.923 T_{\rm obj}  \si{\degreeCelsius}$ according to Fig.~\ref{figS-Tscalib}.
The temperature probe was read using a high-precision handheld thermometer (WIKA, CTH7000, temperature precision: $\pm$ 0.02 K). 
During the calibration measurement, $T_{\rm air}$ was monitored using a data logger (T\&D, TR-72wb, temperature precision: $\pm$ 0.5 K), and we confirmed that $T_{\rm air}$ fluctuates within only $\pm$ 0.5 K over 12 h.
Note that $T_{\rm obj}$ was monitored directly on top of the foil heater.
The temperature stability in the incubator is $\pm 0.2 \si{\degreeCelsius}$ over 12 h when measured by the above temperature probe (see the 0.2-$\si{\degreeCelsius}$ periodic oscillations in Fig.~\ref{figS-Tscalib} (a)).

\begin{figure}[t!]
 \centering
 \includegraphics[scale=1]{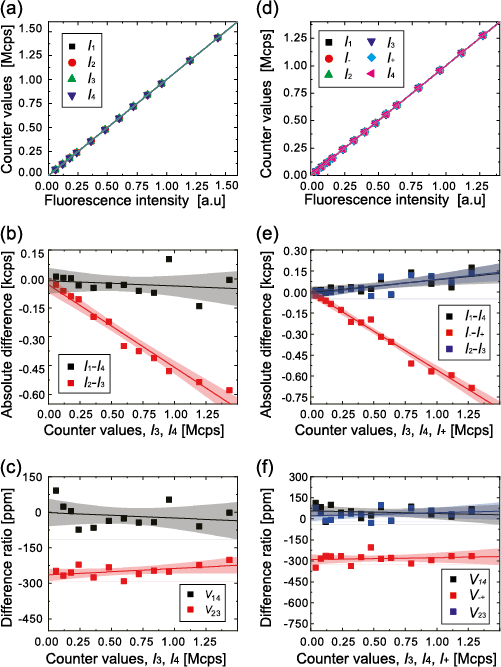}
 \caption{Variation of counter responsivity to the constant TTL pulses from the function generator for 4- and 6-point measurements. (a, d) Total pulse counts of the four counters as a function of input pulse counts for the 4- and 6-point measurements. (b, e) The difference between the counter values for the 4- and 6-point measurements. 
 The solid lines and shaded area represent second-order polynomial fits to the data and their 95-\% confidence intervals, respectively.
 (c, f) The difference in ratios for the 4- and 6-point measurements.
 The solid lines and shaded area represent linear fits to the data and their 95-\% confidence intervals, respectively.}
 \label{fig-gen}
\end{figure}

\section{Results and Discussions} 
\subsection{Optical power dependent instrumental errors of the quantum thermometry}
\label{sec3A}
Theoretically, the counter values on each side of the CW-ODMR spectrum should show the same dependency on the fluorescence counts; however, they show a slightly different dependency.
Figure~\ref{fig3}(a) shows the dependence of the counter values as a function of the fluorescence intensity for the 4-point measurement, where the fluorescence intensity was varied by controlling the laser excitation power using a variable ND filter (specifically VND-1 in Fig.~\ref{fig1}(a)). 
First, the counter values of $I_1$ and $I_4$ are always larger than those of $I_2$ and $I_3$ because they exhibit the difference of the ODMR contrast.
Second, these two sets of the two counters (namely, $I_1$, $I_4$, and $I_2$, $I_3$) exhibit a similar linear increase; however, there are small differences in the values in the order of $10^{-3}$.
Figures~\ref{fig3}(b) and (c), respectively, show the differences of the two counter values ($I_1-I_4$, $I_2-I_3$) and their ratio to the absolute counts that are defined as 
\begin{equation}
V_{ij} = \frac{I_i - I_j}{I_i + I_j}, 
\label{eq4}
\end{equation}
where $i$ and $j$ denote the counter identification numbers.  
The absolute differences are well fitted with the second-order polynomials, and $V_{ij}$ exhibit a linear variation.
The observed small variations of $V_{ij}$ of approximately 0.5 \% is significant in the temperature estimation process, which corresponds to a temperature variation of about a few K.
For the 3-point measurement, the photo-responsivity difference of $I_- - I_+$ is similar to that of the 4-point measurement, while the dynamic range of $I_+$ is relatively large (Fig.~\ref{fig3}(e)), thus deteriorating the curve-fitting precision.
$V_{-+}$ shows the same dependency as $V_{14}$ and $V_{23}$ (Fig.~\ref{fig3}(f)).
For the 6-point measurement, the variations of $V_{ij}$ increase upto $\sim$ 2.0 \%, close to a four-fold increase from those of the 4- and 3-point measurements, as shown in Figs.~\ref{fig3} (g)--(i). 
In particular, the two sets of the differences ($V_{14}$, $V_{-+}$) show large variations (and similar to each other) compared to $V_{23}$ that still shows a similar variation as that of the 4-point measurement. 
As described below, this noticeable difference is a result of distributing the pulse counters in the two DAQ systems and machine-dependent factors such as different clock speeds of DAQ (DAQ-1 and DAQ-2 are in different product lines and use different clock speeds for counting; 100 MHz and 80 MHz for DAQ-1 and DAQ-2, respectively).

We, therefore, tested the behavior of the counter values for the constant TTL pulse trains generated by a frequency generator (Stanford Research Systems, DS345; frequency accuracy, $\pm 5$ ppm). 
The output from the frequency generator was connected to the two DAQ-boards instead of the APD, as shown in Fig.~\ref{fig1}(a). 
Figure~\ref{fig-gen} shows the pulse-responsivity of the counter values, their differences, and $V_{ij}$ for the 4-point and 6-point measurements. 
Note that the photo-responsivity of the 4-point and 3-point measurements are essentially the same and we only compare the 4-point and 6-point measurements.
As expected, all the counter values show almost the same linear increases (Figs.~\ref{fig-gen}(a) and (d)) in contrast to the real photon counting of ODMR, as in Figs.~\ref{fig3} (a), (g). 
The slope of the linear increase is 1.005 for all the counters, which is slightly different from the ideal value of 1. 
This difference is, however, not critical in the multipoint ODMR measurements because the uniform effect of all counters is finally canceled as in Eqs.~\ref{eq:4pnt}--\ref{eq:6pnt-2}. 
Instead, the difference between the counter values in Figs.~\ref{fig-gen}(b) and (h) are one or two orders of magnitude smaller than that of the real photon counting shown in Figs.~\ref{fig3}(b) and (e) (0--300 ppm of $V_{ij}$ compared to 0--1.8 \% of the photon counting case).
Furthermore, $V_{ij}$ is flat over the entire range with an offset and distribution of $<$300 ppm and $\pm$ 50 ppm, respectively, and does not show a significant dependence on the counter values, which can be basically explained by the timing accuracy of the DAQ boards of 50 ppm based on the manufacturer's specification sheet.

\begin{figure*}[t!]
 \centering
 \includegraphics[scale=1.0]{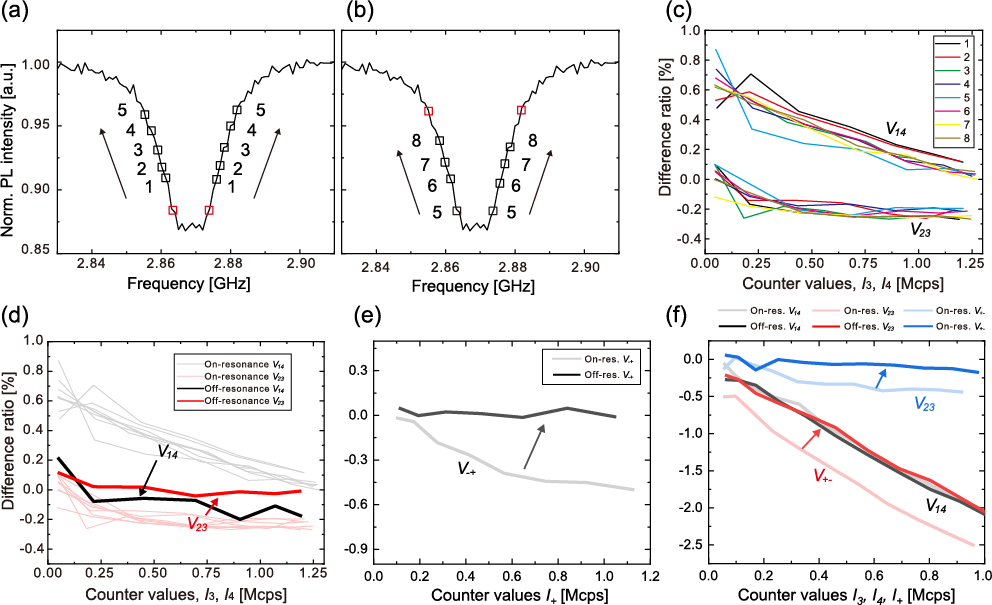}
 \caption{Dependence of the variation of the photo-responsivity on the selection of the microwave frequencies. (a) CW-ODMR spectrum with four selected frequencies; the two bottom frequencies (red) were fixed and the two upper ones (black) are varied to get apart from the bottom two in order of $\bm{1} \to \bm{5}$. (b) CW-ODMR spectrum with four selected frequencies; the two top frequencies were fixed (red) and the two lower ones (black) were varied to approach the top two in order of $\bm{5} \to \bm{8}$. (c) The difference ratio of $V_{14}$ and $V_{23}$ for the eight patterns of the 4-point measurements.
 (d) The comparison between the on-resonant eight patterns of $V_{14}$, $V_{23}$, and their off-resonance profiles exciting around 2.65 GHz for the 4-point measurement. (e) The comparison between the on-resonant $V_{-+}$ and its off-resonance for the 3-point measurement. (f) The comparison between the on-resonant $V_{14}$, $V_{23}$, and $V_{-+}$ and their off-resonance for the 6-point measurement.}
 \label{fig-Fselect}
\end{figure*}

\begin{figure}[th!]
 \centering
 \includegraphics[scale=1.0]{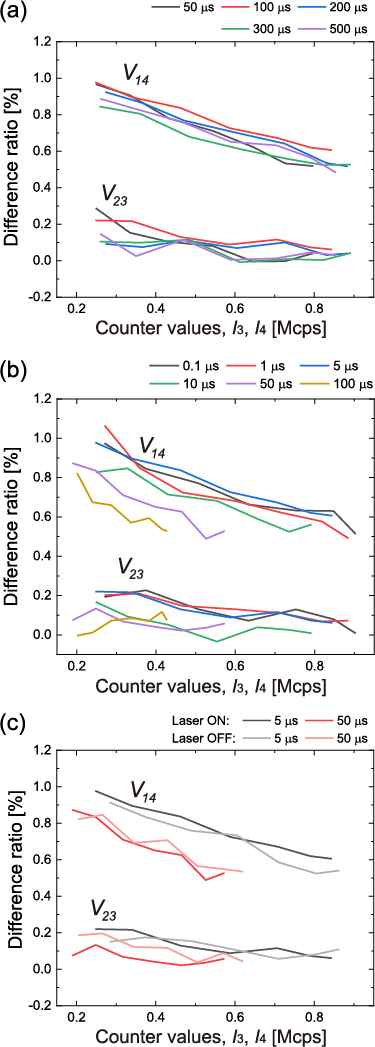}
 \caption{Dependence of difference ratio, $V_{14}$, $V_{23}$, on the photon-counting measurement time, $t_M$ (a) and interval time, $t_{\rm int}$ (b). (c) The difference ratio with $t_{\rm int} = 5$ and 50 $\si{\us}$ for the cases when laser is ON or OFF during the time interval.}
 \label{figS-meas-int-time}
\end{figure}

After quantifying the hardware instrumental error related to the photo-responsivity difference, 
we next check the effect of microwave frequency.
Figures~\ref{fig-Fselect}(a), (b) show eight patterns of choosing frequencies in the 4-point method. 
In Fig.~\ref{fig-Fselect}(a), the upper two frequency points were shifted up, away from the bottom two fixed points, thus increasing the difference of their ODMR depth in the order of $\bm{1} \to \bm{5}$.
On the contrary, in Fig.~\ref{fig-Fselect}(b), the bottom two frequency points were shifted up toward the top two fixed points ($\bm{5} \to \bm{8}$). 
With these eight patterns of frequency selections in the 4-point method, we analyze the behavior of difference ratio of $V_{ij}$.
Figure~\ref{fig-Fselect}(c) shows $V_{14}$ and $V_{23}$ for the respective eight patterns. 
The $V_{ij}$ curves are basically not affected by the frequency selection within the error of $V_{ij}$.
In contrast, Fig.~\ref{fig-Fselect}(d) shows the comparison of $V_{ij}$ between these eight patterns and the off-resonant pattern in which the 4-point measurement was performed by simply shifting the pattern \textbf{3} by 200 MHz to the lower frequency side, namely, around 2.65 GHz. 
The $V_{ij}$ values now turn small within the range of $\pm$ 0.1 \%. 
While this value is still one order of magnitude larger than the case of Fig.~\ref{fig-gen}, the $V_{ij}$ shows the same flat photo-responsivity as that of the counter responsivity (Fig.~\ref{fig-gen}(c)).
As shown in Fig.~\ref{fig-Fselect}(e), this on/off-resonance behavior of the 4-point measurement is also observed in the 3-point measurement.
In the 6-point measurement, shown in Fig.~\ref{fig-Fselect}(f), $V_{23}$ exhibits the same on/off-resonance behavior as $V_{23}$ of the 4-point measurement, 
whereas the difference between $V_{14}$ and $V_{-+}$ vanish, exactly matching each other. 
This fact indicates that the use of different DAQ boards created the same dependency for $V_{14}$ and $V_{-+}$, but the difference between $V_{14}$ and $V_{-+}$ that were observed in the ODMR measurement in Fig.~\ref{fig3}(i) should be related to the NV spin resonance itself.
Note that the we have confirmed that the photo-responsivity difference is also observed in the permuted 4-point method that has been recently reported~\cite{choi2019}.

We further confirmed that the irradiation of laser and microwave does not cause this photo-responsivity difference by measuring its dependence on the duration of the photon-counting ($t_M$) and interval times ($t_{\rm int}$). 
Figure~\ref{figS-meas-int-time} (a) shows the dependence of the difference ratio acquired with the 4-point method on the measurement time, $t_M$.
The dependence of $V_{14}$ and $V_{23}$ does not change significantly when the measurement time is increased from 50 to 500 $\si{\us}$. 
Similarly, Fig.~\ref{figS-meas-int-time} (b) shows the dependence on the interval time ($t_{\rm int}$) when $t_{\rm int}$ is increased from 0.1 to 100 $\si{\us}$. 
As the interval time increases, $V_{14}$ is shifted to the left (particularly 50 and 100 $\si{\us}$) because of the reduction of the detected photon counts per second. 
However, the overall change of difference ratio between the minimum and maximum counter values does not change because the optical power itself is kept at the same level.
Furthermore, the laser irradiation during the interval time does not affect the photo-responsivity difference as shown in Fig.~\ref{figS-meas-int-time} (c).

Summarizing these experimental results, the random nature of photon-counting events just increases the measurement accuracy of DAQ boards, and it does not cause photo-responsivity difference such as linear dependency of $V_{ij}$.  
Although using different DAQ counting boards may cause the photo-responsivity difference, it can be identified by checking the photo-responsivity difference between the on/off-resonance behaviors.
The final remaining photo-responsivity difference should be related to the intrinsic nature of the ODMR signal. 
While the mechanism of this ODMR behavior cannot be explained well presently, it may be related to some optical-power-dependent spectral distortion of the ODMR dip.
For example, it is well known that the ODMR depth and linewidth depend on the laser excitation intensity~\cite{PhysRevB.84.195204}.
Such optical power dependency and some other associated minor phenomena may cause the present photo-responsivity difference.
It should be noted that our CW-ODMR measurements were not able to address this small asymmetry because of the limitations of acquiring sufficiently high signal-to-noise ratio data. 
In our standard configuration, the whole spectral measurement accumulates 20 Mcts  of photons at each point of frequency, which gives 4.5 kcts as a photon shot noise. 
The ODMR contrast is approximately 10\%, i.e., the ODMR depth is 2 Mcts of contrast.
The observed spectral asymmetry is therefore to be 2 kcts (0.1 \%), which needs high degree of data accumulation. 
Furthermore, the long integration time of the whole spectral measurement (2--20 min depending on the photon count) inevitably incorporates low-frequency noise, which affects the spectral shape.
It is also necessary to work on NV centers in bulk diamonds  to account for the underlying material physics of NV centers.
We also note that the impedance mismatch between the APD output (50 $\Omega$) and the DAQ inputs (1 k$\Omega$) is not likely to affect the current observation.
Impedance mismatch causes reflections in the coaxial electrical line and may affect the detected pulse numbers. 
However, the pulse counter experiments using the frequency generator (Fig.~\ref{fig-gen}) also have an output impedance of 50 $\Omega$, the same as that of APD. 
If the impedance mismatch causes the observed photo-responsivity difference, it should have been observed in Fig.~\ref{fig-gen}.

\subsection{Practical approach for canceling the instrumental errors}
Whereas the exact origin of the variations of the counter responsivity is not completely interpreted, these effects can be practically eliminated by subtracting pre-known systematic errors from the original counter values. 
Such a error-correction filter is necessary particularly for the measurements operated in the low-photon regime ($I_{\rm tot} <$ 1 Mcps) because it can create significant artifacts in the frequency-shift estimate (i.e. about 300 kHz corresponding to several degrees).
While conducting experiments, the counter responsivity is measured each time before performing the multipoint ODMR measurements.
After acquiring the counter responsivity data as presented in Fig.~\ref{fig3}, the data were fitted with second-order polynomials.
With the fitting parameters of the second-order polynomials, the original counter values can be corrected as follows: 
\begin{equation}
\begin{split}
I_3^{\rm C} &= I_3^{\rm NC} + [ a_0 + a_1 I_3^{\rm NC} + a_2 (I_3^{\rm NC})^2], \\
I_4^{\rm C} &= I_4^{\rm NC} + [ b_0 + b_1 I_4^{\rm NC} + b_2 (I_4^{\rm NC})^2], \\
I_+^{\rm C} &= I_+^{\rm NC} + [ c_0 + c_1 I_+^{\rm NC} + c_2 (I_+^{\rm NC})^2], \\
\end{split}
\label{eq5}  
\end{equation}
where $I_i^{\rm C}$, $I_i^{\rm NC}$, $a_k$, $b_k$, and $c_k$ denote the error-corrected photon counts, original photon counts (no error correction), and coefficients of the second-order polynomials for $I_3$, $I_4$, $I_+$, respectively.
Table~\ref{tbl0} summarizes these fitting parameters used in Fig.~\ref{fig3}. 
In this way, we can eliminate the systematic errors of the measurement systems included in the original counter values.
Note that this error correction does not explicitly depend on $I_1$, $I_2$, or $I_-$; the data points located on the left side of the ODMR spectrum are used concomitantly for the calculation in Eqs.~\ref{eq:4pnt}--\ref{eq:6pnt-2}.
This independence of the other side of the ODMR spectrum is important to isolate the error correction from the temperature derived ODMR shift.

\begin{figure*}[th!]
 \centering
 \includegraphics[scale=1.0]{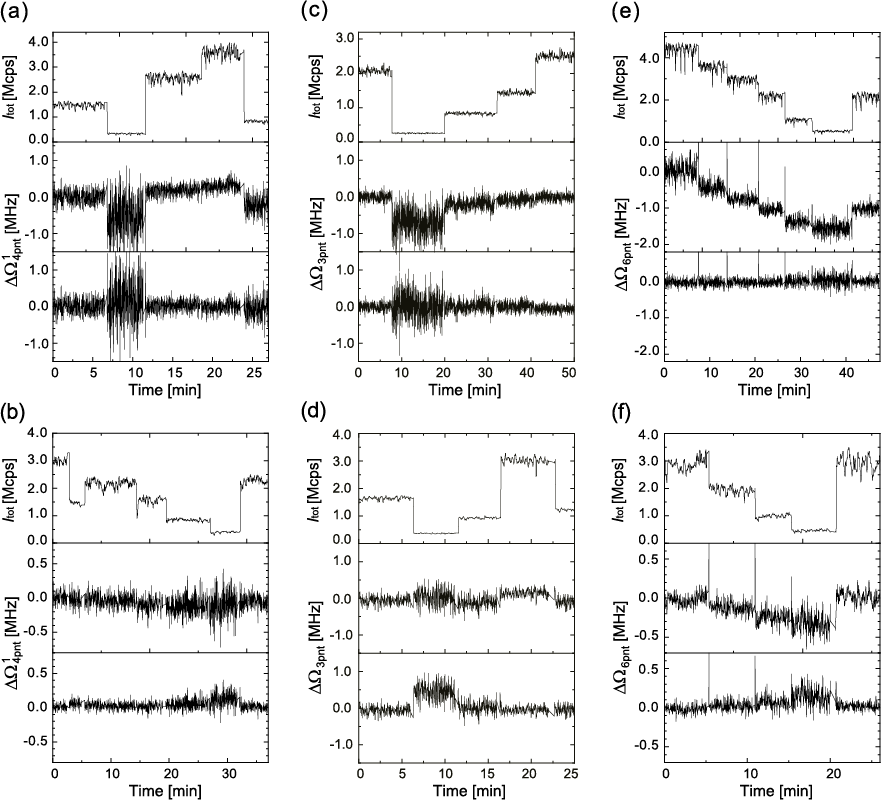}
 \caption{Time profiles of $I_{\rm tot}$ (top), $\Delta \Omega _{\rm 4pnt}^1$ without (middle) and with the error correction (bottom) in the 4-point measurement for the two cases of the laser intensity variation (a) and the PL intensity variation (b). Time profiles of $I_{\rm tot}$ (top), $\Delta \Omega _{\rm 3pnt}$ without (middle) and with the error correction (bottom) in the 3-point measurement for the cases of the laser intensity variation (c) and the PL intensity variation, respectively (d). 
 Time profiles of $I_{\rm tot}$ (top), $\Delta \Omega _{\rm 6pnt}$ without (middle) and with the error correction (bottom) in the 6-point measurement for the cases of the laser intensity variation (e) and the PL intensity variation (f).}
 \label{fig5-8}
\end{figure*}

\begin{figure*}[th!]
 \centering
 \includegraphics[scale=1.0]{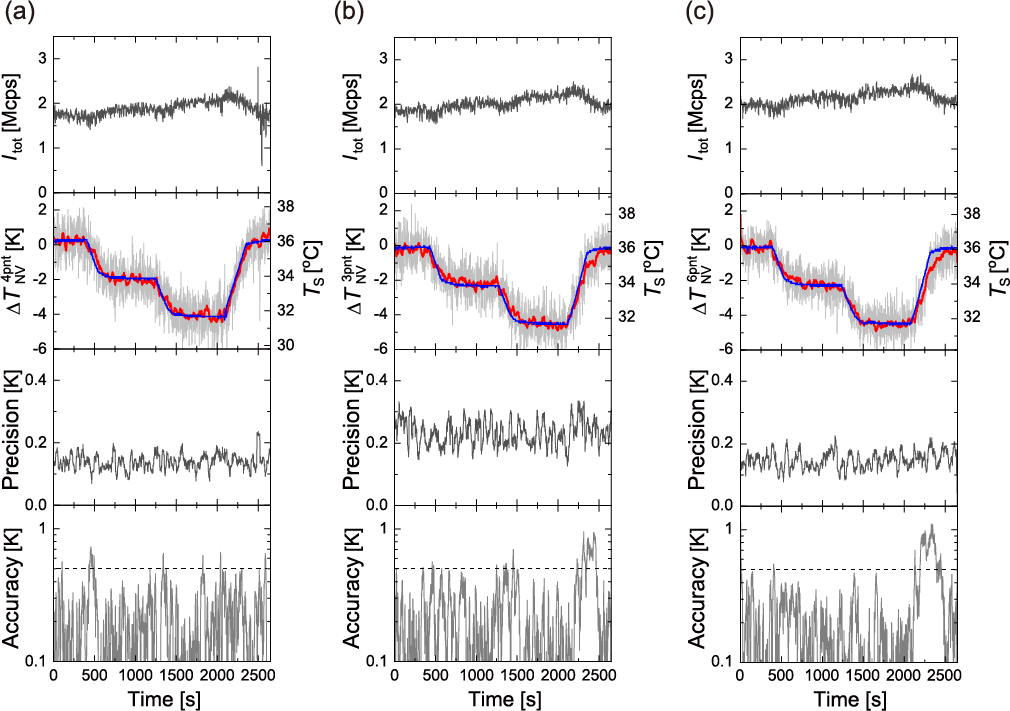}
 \caption{Time profiles of $I_{\rm tot}$ (top), $\Delta T_{\rm NV}$ (second top), precision (second bottom), and accuracy (bottom) for the 4-point (a), 3-point (b), and 6-point measurements (c), measured on the same ND sequentially. In the second top panel, the gray color represents the 1-s sampling data, red represents 20-point moving average, and blue represents the sample temperature $T_{\rm S}$.}
 \label{fig-comp}
\end{figure*}

Figure~\ref{fig5-8}(a) shows $I_{\rm tot}$, $\Delta \Omega _{\rm 4pnt}^1$ without and with the error correction for the 4-point measurement, respectively, when the laser intensity is intentionally varied.
The error correction suppresses the signal drift. 
While the temperature estimate shows an offset drift when the laser intensity is changed without the error correction, it no longer shows the drift when the error is corrected.
It also works well when the fluorescence intensity is varied for a constant laser intensity, as can be seen in Fig.~\ref{fig-comp}(b); specifically a variable ND filter (VND-2 in Fig.~\ref{fig1}(a)) was varied to directly control the fluorescence intensity detected by the APD. 
Figures~\ref{fig5-8}(c)--(f) show the corresponding data for the 3- and 6-point measurements when either of the laser intensity or fluorescence intensity is changed.
The error correction is still effective for the 3- and 6-point measurements.
In all the cases, the noise is generally reduced as a result of the error correction in addition to the drift cancellation.
Note that the error correction does not work very well in the low-photon regime $I_{\rm tot} < 0.5$ Mcps owing to the inaccuracy of the fitting parameters; 
It is, therefore, necessary to always perform measurements at $I_{\rm tot} > 0.5$ Mcps  to ensure that drift is negligible.
Note also that the fitting errors are propagated to the absolute accuracy of the temperature measurement (see Eqs.~\ref{eq:prop4pnt}--\ref{eq:prop6pnt} for the error-propagation equation). 
Figure~\ref{fig-errorprop} shows the dependence of the propagated errors on the fluorescence intensity for the present three types of multipoint ODMR methods.
In the 4- and 3-point methods, the errors of the absolute measurement accuracy is 2-3 K for the most of the photon count range and can be increased up to 6 K when the photon counts are decreased.
In the 6-point method, the accuracy is increased to 6-12 K because of the large instrumental errors of DAQ devices.
Note that this inaccuracy means there is a constant uncertain offset in the measured $\Delta T_{\rm NV}$ because we do not change the error-correction parameters during the temperature measurement.

\renewcommand{\arraystretch}{1.2}
\begin{table*}[th!] 
  \caption{Fitting parameters of the second order polynomials}
  \label{tbl0}
  \centering
  \begin{tabular}{p{20mm}p{40mm}p{40mm}p{40mm}}
    \hline \hline
	Parameters & 4-point & 3-point & 6-point\\
    \hline \hline
	$a_0 +\delta a_0$ & (-0.11 $\pm$ 1.34) $\times 10^{2}$ &--   & (0.27 $\pm$ 3.57) $\times 10^{2}$ \\
	$a_1 +\delta a_1$ & (1.42 $\pm$ 0.15) $\times 10^{-2}$ & --  & (-2.23 $\pm$ 2.09) $\times 10^{-3}$\\
	$a_2 + \delta a_2$ & (-2.12 $\pm$ 0.33) $\times 10^{-8}$ & --  & (-4.45 $\pm$ 0.26) $\times 10^{-8}$\\
	\hline
	$b_0 + \delta b_0$ &  (2.30 $\pm$ 2.33) $\times 10^{2}$  & -- & (1.80 $\pm$ 5.19) $\times 10^{2}$ \\
	$b_1 + \delta b_1$ &  (-0.92 $\pm$ 2.68) $\times 10^{-3}$  & --  & (-1.58 $\pm$ 0.31) $\times 10^{-2}$\\
	$b_2 + \delta b_2$ &  (-1.46 $\pm$ 0.61) $\times 10^{-8}$  & --  & (-3.08 $\pm$ 0.38) $\times 10^{-8}$\\
	\hline
	$c_0 + \delta c_0$ & -- & (1.05 $\pm$ 0.45) $\times 10^{3}$  & (4.77 $\pm$ 7.10) $\times 10^{2}$ \\
	$c_1 + \delta c_1$ & -- & (-7.19 $\pm$ 2.18) $\times 10^{-3}$ & (-1.07 $\pm$ 0.44) $\times 10^{-2}$ \\
	$c_2 + \delta c_2$ & -- & (-3.47 $\pm$ 2.11) $\times 10^{-9}$  & (8.69 $\pm$ 5.64) $\times 10^{-9}$ \\
	\hline
  \end{tabular}
\end{table*} 
\renewcommand{\arraystretch}{1.0}

\subsection{Measuring stepwise temperature change using the multi-point methods}
Using these error corrections, we can now measure the temperature change of NDs while the photon counts are dynamically changed, similar to the heating/cooling of the oil-immersion objective, which has been technically challenging because the change of objective temperature strongly shifts the focal position and this positional shift causes photon count noise.

Figures~\ref{fig-comp}(a)--(c) show the time profiles of the temperature estimates for the three types of the multipoint ODMR measurements by working on the \textit{same} single ND with the constant optical and microwave power in the 4-, 3-, and 6-point methods, respectively.
Each measurement takes approximately 40 min and was performed sequentially, thereby taking $\sim 2$ h throughout these three measurements.
As the temperature of the sample ($T_{\rm S}$) is decreased stepwise by $\sim 2$ K,
$I_{\rm tot}$ increases due to the improvement of the fluorescence quantum efficiency, as previously reported~\cite{plakhotnik2010luminescence} (top panel). 
As shown in the second top panels, all multipoint ODMR methods successfully follow the stepwise temperature change. 
Interestingly, the scaling factor between $T_{\rm S}$ and $\Delta T_{\rm NV}^{i {\rm pnt}}$ is different among the three methods, giving different temperature dependency of zero-field splitting $\alpha$ ($-54.1$, $-95.0$, $-67.5 \ \si{\kHz} \cdot \rm{K}^{-1}$ for the 4-, 3-, and 6-point measurements, respectively), while working on the same single ND. 
The difference between the 4- and 6-point measurements originates from the fact that the $\Delta \Omega _{\rm NV}^{\rm 6pnt-3}$ (probing the deepest region of the dip) is largely deviated from the other measurements of $\Delta \Omega _{\rm NV}^{\rm 6pnt-1}$ and $\Delta \Omega _{\rm NV}^{\rm 6pnt-2}$. 
This deviation indicates that the near-dip region shifted more than the base region above the FWHM, thereby revealing a slight distortion of spectral shape under the temperature change.
The relatively large $\alpha$ of the 3-point method is related to the assumption of single Lorentzian shape for the ODMR spectrum.
In most cases, the ODMR spectrum cannot be simplified as a single Lorentzian and there must be a deviation from the Lorentzian-based estimation in the real measurements.
As described in Fig.~\ref{fig-3pnt-theory}, Eq.~\ref{eq:3pnt} holds only when the ODMR spectrum exactly matches single Lorentzian. 
It otherwise provides a significant deviation of the estimation from the real temperature change by a factor of upto 2 depending on the real profile. 
Note that, in our experiment, the observed values of $\alpha$ (particularly of 4- and 6-point methods) are generally smaller than previous reports of -74 $\si{\kHz} \cdot \rm{K}^{-1}$ probably because of the heating method in which NDs are heated from the bottom and cooled by the surrounding air. 
There is also a material inhomogeneity of $\alpha$ both for bulk diamonds and NDs~\cite{PhysRevLett.104.070801,foy2019wide,fujiwara2019realtime}.
However, the present comparison between the multi-point ODMR methods is not affected by the difference of the absolute values.

The precision and accuracy of the NV thermometry during the measurement are also shown in the two bottom panels in Fig.~\ref{fig-comp}.
The 4- and 6-point methods give almost the same precision (0.14 and 0.15 K, respectively), but the 3-point method gives a slightly large error (0.23 K). %
The sensitivity is then calculated as 0.9, 1.4, and 0.9 $\rm{K} /\sqrt{\si{Hz}}$ for the 4-, 3-, and 6-point methods, respectively, for this particular ND. 
The accuracy is not affected by the measurement methods and is $<$ 0.5 K, which is common to all the methods.
It should be noted that the present accuracy is overestimated because $\Delta T_{\rm NV}$ in the stable states is significantly influenced by the environmental temperature fluctuation ($T_{\rm air}$).
The potential accuracy of the NV thermometry should be better than the present value.

The presently used moving average of $\Delta T_{\rm NV}^{\rm 4pnt}$ over 20 data points can track the dynamic change of the temperature. The selection of the averaging range needs careful consideration on the sensor noise (as described in the following section of Allan variance analysis) and required time resolution.
While the 20-point moving average is best for the present data, one could use other filtering techniques for some applications to achieve temporal resolution and temperature precision simultaneously, such as Kalman filter (see Fig.~\ref{fig-kalman}), because these filtering techniques can work more efficiently in some situations such as transient dynamics with noisy measurements~\cite{7516277}.

\begin{figure*}[th!]
 \centering
 \includegraphics[scale=1.1]{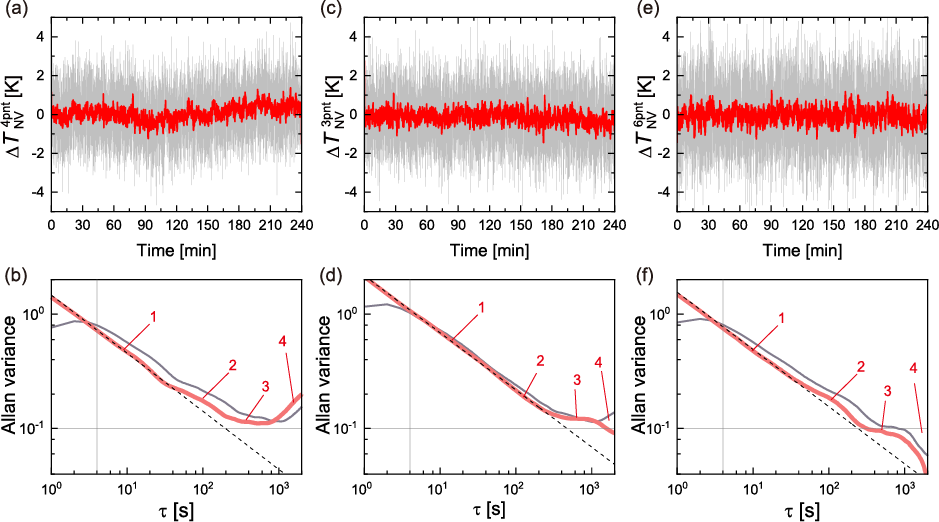}
 \caption{Time profiles of $\Delta T_{\rm NV}$ for the 4-point (a), 3-point (c), and 6-point measurements (e). The corresponding Allan variance curves for the 4-point (b), 3-point (d), and 6-point measurements (f). The thick red line shows the Allan variance assuming the temperature is measured every 1 s, regardless of the tracking time. The gray line indicates the Allan variance in the real-time scale, including the tracking time (the profile is obtained by interpolating the original data). The vertical straight line at $\tau = 4$ s indicates the tracking period. The black dotted lines indicate the linear line with the slope of -0.5. }
 \label{fig-alavar}
\end{figure*}

\subsection{Noise analysis of the ND quantum thermometry}
We next analyze the noise profiles of the thermometry by recording the temperature time profiles for a long term in the respective multi-point ODMR methods. 
Figures~\ref{fig-alavar}(a), (b) show the time profiles of $\Delta T_{\rm NV}^{\rm 4pnt}$ over 4 h and its Allan variance for the 4-point measurement, respectively.
This measurement was performed at a constant temperature of $T_{\rm obj}$ = 37 $\si{\degreeCelsius}$ ($T_S = 36 \si{\degreeCelsius}$).
The Allan variance profile shows four regions; it first shows a linear decrease with the slope of $-0.5$ until 40 s, which indicates white noise (Region 1).
Between 40 and 200 s, the slope of the linear decrease turns by a relatively smaller value of $-0.35$ (Region 2, see Fig.~\ref{figS-alavarslopes}).
The profile then lands on a plateau lasting from 200 to 800 s (Region 3), followed by a linear increase beyond 800 s (Region 4). 
Region 3 and 4 are mostly related to the temperature instability of the environment. 
Region 3 exhibits the temperature instability of the incubation chamber for the time scale of minutes. 
Region 4 corresponds to the air-conditioned room temperature fluctuation with a period of $\sim$ 20 min (see Fig.~\ref{figS-Tscalib}). 
Region 4 sometimes includes instability of NV spin properties; 
for example, Fig.~\ref{fig-alavar}(a) shows a bumpy profile while there were no noticeable changes in the room temperature. 
Such long-term fluctuations are sometimes observed in our experiments of the ND quantum thermometry, which may be related to the NV stability under the optical excitation.
At present, the noise source of Region 2 has not been clarified, but it may be related to some mechanical instability of the confocal microscope.
For these reasons, the time window of the moving average filter is optimal at around 40 s, corresponding to 20 adjacent points.

Figures~\ref{fig-alavar}(c)--(f) show the corresponding long-term profiles of $\Delta T_{\rm NV}$ and their Allan variance profiles for the 3-point and 6-point measurements.
Both of them show the common four regions in the Allan variance profiles, while there are slight differences in the prominence of Region 2 and 4.
Note that Region 4 in the 6-point measurement further goes lower in this figure; however, it increases again beyond 1500 s, as shown in Fig.~\ref{figS-alavar6pnt}(c). 
Note also that the first plateau of noise profile until $\sim$ 4 s is due to the uneven time spacing of the data in the Allan variance calculation or interpolation effect as described in Appendix~\ref{sec:ApH}.

\subsection{Artifact analysis for the temperature estimation}
We have so far analyzed the instrumental errors of our thermometry system when estimating the temperature based on the limited available information of the ODMR spectrum. 
Our study highlights the importance of analyzing all possible factors that may affect the frequency shift estimation. 
Signals from the NV quantum thermometers may be more complicated in biochemical environments including nanofiber structures~\cite{price2019quantum}, lipids~\cite{Kaufmann10894,sotoma2018highly}, cells~\cite{simpson2017non,kucsko2013nanometre,claveau2018fluorescent}, and living animals~\cite{choi2019,fujiwara2019realtime,mohan2010vivo,Simpson:14,lin2016nanodiamond}, 
which inevitably increases the possibility of sensory artifacts of NV centers.
Table~\ref{tbl1} summarizes such factors that may cause frequency shift, linewidth change, and fluorescence intensity, which ultimately can appear as artifacts in the multipoint ODMR measurements. 

The most direct factors are the magnetic field, stress, temperature, and electric field, which have been major sensing targets of diamond NV sensors themselves~\cite{schirhagl2014nitrogen}.
The magnetic field is the most prominent factor that causes a frequency shift via the Zeeman effect. 
Given that the effect of the static magnetic field is canceled in the present method (Eqs.~\ref{eq:4pnt} and \ref{eq:6pnt-1}), slowly-varying magnetic field, including the geomagnetic field, does not seem to induce artifacts as long as the magnetic field is static for more than 1.0 s.
In this context, the application of quantum thermometry to brain activity requires great care during data analysis because neural activity generates strong and complicated magnetic fields for the short time-scale via the internal electric current, which is used for magnetoencephalography (MEG) ~\cite{barry2016optical,caruso2017vivo}. 
Conversely, the magnetometry applications of NV centers for MEG  need to take into account the temperature variation in brain tissue structures.
Stress is not likely to affect the ODMR measurements in most biological experiments because biological pressures are considerably small and do not affect the crystal field of NV centers in diamonds (15 kHz/MPa~\cite{doherty2014electronic,cai2014hybrid}).
An electric field can also cause a frequency shift by inducing stark shift~\cite{dolde2011electric,iwasaki2017direct}.
While there is a proposal to measure local electric fields in a biological context such as a trans-membrane electric field generated by the cell-membrane potential~\cite{schirhagl2014nitrogen}, a more critical situation would be when bio-molecules possessing large electron affinity are adsorbed on the ND surface, which is known to be adsorbed by various biomolecules in cells or living organisms~\cite{doi:10.1002/anie.201905997,doi:10.1246/cl.141036,lin2015protein}. 
Moreover, this mechanism has been used to manipulate the ND spin properties by surface functionalization~\cite{zhang2018hybrid,raabova2019diamond}.
The surface functionalization is thus important to control both the sensitivity and robustness of NV spins systems.

Other factors that are particular to biological environments include pH, ionic strength, and water adsorption to the ND surface. %
Previous experiments based on the CW-ODMR spectral measurements have confirmed that these factors do not change both the ODMR frequency and the spectral shape in a wide range of pH and various ionic solutions~\cite{fujiwara2019monitoring,sekiguchi2018fluorescent}. 
Because the precision of the CW-ODMR-based methods used in the literature was only in the range 1-2 $\si{\degreeCelsius}$, small effect within this precision range may be detected in the multipoint ODMR methods. 
It is also important to consider the effect of pH on the emission stability of negatively charged NV centers~\cite{karaveli2016modulation}, as the emission instability causes variations in the fluorescence intensity. 
In addition, NDs move randomly in cells and worms by Brownian motion, intracellular transportation, and body motions~\cite{Simpson:14,choi2019,fujiwara2019realtime}. 
These particle motions cause photon count fluctuations that may affect the multipoint ODMR measurements if the instrumental errors are not fully removed.
Microwave irradiation may cause a change in the temperature during the measurement process due to microwave water heating that causes ODMR shift~\cite{woo2000differential}.
Because microwaves are always irradiated during the measurements  and would not affect the relative temperature measurements, local heating due to some conformational changes of biological samples may change the local dielectric permeability, thereby affecting the local heat-generation rate.
Such artifacts also need to be seriously considered if the observed temperature signal from the NV centers is not very straightforward.

One possible approach to validate the measured temperature data is comparing/combining the ODMR thermometry with all-optical NV thermometry~\cite{Plakhotnik2014all,Plakhotnik_2015,fukami2019all-optical,plakhotnik2010luminescence} and other optical nanothermometry techniques~\cite{okabe2012intracellular,doi:10.1002/anie.201915846,tsuji2017difference,qiu2020ratiometric,jaque2012luminescence,del2018vivo}. 
Such dual or multi-modal temperature measurements is meaningful particularly for biosensing applications because of the complex factors that might cause artifacts of the ODMR shift.

\renewcommand{\arraystretch}{1.2}
\begin{table*}[t!] 
  \caption{Real and artifact factors that can cause ODMR shift of NV centers. 
  $^\ast$: transnational Brownian motion. $^\dagger$: rotational Brownian motion.}
  \label{tbl1}
  \centering
  \begin{tabular}{p{30mm}p{30mm}p{20mm}p{20mm}p{20mm}p{20mm}p{30mm}}
    \hline \hline
	\multirow{2}{*}{Source} & Physical mechanism  & ODMR & \multirow{2}{*}{Linewidth} & Photon & \multirow{2}{*}{Ref.}\\
	 & (Artifact source)  & shift &  & count  & \\
    \hline \hline
	Magnetic field	 & Zeeman & Yes & No & No &  \cite{maze2008nanoscale}\\
	\hline
	Stress & Jahn-Teller & Yes & No & No &  \cite{cai2014hybrid}\\
		\hline
	Temperature & Lattice-expansion (stress) & Yes & No & No &  \multirow{2}{*}{\cite{kucsko2013nanometre,PhysRevLett.104.070801}}\\
		\hline
	Electric field & Stark &	Yes	 & No	& No &  \cite{dolde2011electric}\\
	\hline
	pH, ions, 	 & Charge state & Maybe & Maybe & \multirow{2}{*}{Yes} &  \multirow{2}{*}{\cite{fujiwara2019monitoring,sekiguchi2018fluorescent,fujisaku2019ph,sigaeva2019optical}}\\
	water adsorption & fluctuation & No & No &  & \\
		\hline
    Biomolecule & Various & \multirow{2}{*}{No} & \multirow{2}{*}{No} &  & \multirow{2}{*}{\cite{ermakova2013detection,raabova2019diamond,doi:10.1002/anie.201905997,doi:10.1246/cl.141036,lin2015protein,hsiao2016fluorescent}} \\
    adsorption & factors & & & \\
	\hline
	\multirow{2}{*}{TBM$^\ast$} &	photon count & \multirow{2}{*}{No} & \multirow{2}{*}{?} & \multirow{2}{*}{?} & \multirow{2}{*}{\cite{Simpson:14}}  \\
     & fluctuation & & & & \\
     	\hline
	\multirow{2}{*}{RBM$^\dagger$} & \multirow{2}{*}{Geometric Phase} &		\multirow{2}{*}{No}	& \multirow{2}{*}{Yes} &	\multirow{2}{*}{No} &  \multirow{2}{*}{\cite{fujiwara2018observation}}\\
	 & & & & & \\
	 	\hline
	Microwave heating & Water absorption & Yes & No & Yes &  \cite{woo2000differential}\\
    \hline \hline
  \end{tabular}
\end{table*} 
\renewcommand{\arraystretch}{1.0}

\section{Conclusion and Future Perspectives}
In this study, we reported the experimental details and protocols of real-time multipoint ODMR measurements, i.e., 4-point, 3-point, and 6-point methods. 
The multipoint ODMR measurement is a process used to estimate the frequency shift of ODMR based on limited fluorescence intensity data at several frequency points. 
Therefore, careful analysis is required with respect to the estimation of the frequency shift because unexpected factors might generate artifacts.
The difference in the photo-responsivity between the photon counters is such an example. 
We have shown that this photo-responsivity difference originates from the intrinsic nature of NV spins.
In case of using multiple DAQ boards in the 6-point method, an instrumental error of experimental hardware is further added.
Both these error sources are very small and were not noticed in the previous NV quantum experiments.
A careful analysis of the different types of multi-point ODMR measurements has suggested a complicated spectral distortion when the thermal shift of ODMR dip occurs.
We have proposed a practical method to cancel these artifacts, whereby the artifact values are subtracted from the obtained counter values, based on the pre-characterized photo-responsivity curves.
Using developed real-time thermometry, we succeeded in measuring the temperature of single NDs during stepwise temperature change.
We also quantitatively compared the precision and noise sources of the 4-, 3-, and 6-point measurements. 
We also discussed possible noise sources and artifacts in the quantum thermometry that should be considered in biosensing experiments. 

An important implication of the present study is that quantum sensing requires both, a high sensitivity and the effective implementation of sensors for realistic measurements. 
The present study identifies a variety of artifact sources for real-time operation. 
The combination of multipoint ODMR measurements and particle tracking also requires great care to avoid these artifacts. 
Particle tracking is a feedback process used to maximize the fluorescence counts of NDs that move away from the focus, and can be coupled with variations of fluorescence intensity derived from temperature changes in NV centers.
The present particle tracking is not significantly coupled with the temperature estimation because of the constant re-positioning time. It may be coupled with thermometry when a fast particle-tracking algorithm is employed. 
Comparing or combining the ODMR thermometry with all-optical NV thermometry and other optical nanothermometry techniques will be important to confirm  the results of the measured temperature data particularly for biosensing applications because of the complex factors that might cause artifacts of the ODMR shift.
Such studies on the implementation of quantum sensors into realistic measurement systems are crucial to the future development of quantum sensing.

\section*{Acknowledgment}
The authors thank J. Choi, H. Ishiwata, S. Inouye, E. Kage-Nakadai, N. Komatsu, M. D. Lukin, P. Maurer, M. Turner, J. Twamley, Y. Umehara, R. Walsworth, and H. Zhou for fruitful discussions and technical assistance in experiments.
This work is supported in part by Osaka City University Strategic Research Grant 2017 \& 2018 (M.F., Y.S., A.D.), and JSPS-KAKENHI (M.F.: 16K13646, 17H02741, 19K21935, 20H00335. Y.S.: 19K14636, M.F. and Y.S.: 20H00335).
M.F. acknowledges funding by the MEXT-LEADER program, Sumitomo Research Foundation and Murata Science Foundation.
O.B., A.D. acknowledge funding by the Deutsche Forschungsgemeinschaft DFG (FOR 1493).

\clearpage
\appendix
\renewcommand{\thefigure}{S\arabic{figure}}
\setcounter{figure}{0}  

\section*{Appendix}

\section{Method for the multipoint selection}
\label{A:A}
\begin{figure}[t!]
 \centering
 \includegraphics[width=80mm]{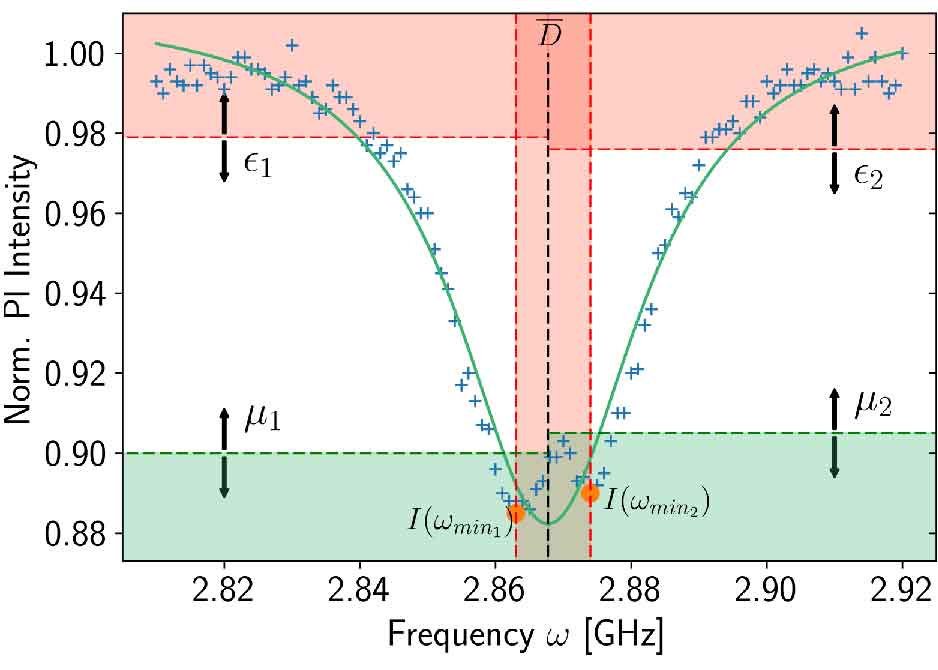}
 \caption{ODMR spectrum showing areas of data points that have been excluded prior to the fitting of the slopes (red areas) for either being part of the baseline noise (regulated via $\epsilon_{1,2}$) or dip noise, i.e., being in between the two local minima, and post the fitting (green area) for being in a considerably close proximity to the dip (regulated via $\mu_{1,2}$). A single Lorentzian dip (green line) indicates the approximate position of the zero-field splitting (dotted black line). }
 \label{fig-fpointselction}
\end{figure}

To determine the six frequency points at which the ODMR spectra are probed, the entire CW-ODMR spectral shape is considered for analysis. 
The analysis is divided into the following chronological steps:

\begin{enumerate}[leftmargin=*]
\item
The ODMR spectra are fitted using a sum of two Lorentzian functions of the form of 
\begin{equation}
\begin{split}
    F_{\rm double}(\omega) =& Y_0 + 2A_1 \left[\frac{\Gamma_1}{4(\omega - \bar{D_1})^2+\Gamma_1^2} \right] \\
    & + 2A_2 \left[\frac{\Gamma_2}{4(\omega - \bar{D_2})^2+\Gamma_2^2} \right]
\end{split}
\end{equation}
to account for the increased dip splitting.
Here, $Y_0$, $A_i$, $\Gamma_i$, $\bar{D_i}$ denote the global offset, spectral area, HWHM, and dip frequency of the $i$-th Lorentzian, respectively.
We however simplified this double Lorentzian using a single Lorentzian as follows: 
\begin{equation}
F_{\rm single}(\omega) = Y_0 + 2A_1 \left[\frac{\Gamma}{4(\omega - D)^2+\Gamma^2} \right].
\end{equation}
The zero-field splitting ($D$) is then approximately given by
\begin{equation}
    \bar{D} \approx D = \bar{D_1} + \frac{\bar{D_2} - \bar{D_1}}{2}.
\end{equation}
This approximation is valid as the difference between $D$ and $\bar{D}$ is one or two orders of magnitude smaller than the linewidth.

\item
Each domain is then checked for a local intensity minimum, and data points to the
right of the local minimum ($\omega > \omega _{\rm min1}$) in the first domain and 
data points to the left of the local minimum ($\omega > \omega _{\rm min1}$) in the second domain are extracted, as indicated in Fig.~\ref{fig-fpointselction}. 
This is an important step in case of dip splitting as it sets the ground for a solid first guess for the linear fit for the two slopes.

\item 
Data points that can be regarded as baseline noise are specified on each
slope. This is done by comparing the absolute ODMR dip intensity value of that data point, i.e., $|I_{\rm dip} (\omega)| = |I (\omega) - Y_0|$ to our preset fraction of the maximum value of the ODMR dip intensity, i.e., $|I_{\rm dip}^{\rm base} (\omega_{\rm min1,2})| = \epsilon_{1,2} |I (\omega_{\rm min1,2}) - Y_0|$ with $0 \leq \epsilon_{1,2} \leq 1$.
In the case that $|I_{\rm dip} (\omega)| < I_{\rm dip}^{\rm base} (\omega_{\rm min1,2})|$, the data point is excluded (Fig.~\ref{fig-fpointselction}).

\item
The two linear slopes of the ODMR spectrum are recognized by linear fits. 
In the first step of each iteration, the routine generates two datasets, of which the first misses the first element and the second misses the last element of the initial dataset. 
Both datasets are then fitted and the resulting residues are compared. 
The dataset with the lower residue is passed on. 
It is in this way, the extent of the linear slope is narrowed down. 
The routine is executed until a preset maximum-allowed residue is surpassed or the amount of elements in the dataset reaches the minimum-allowed amount.

\item
Two functions are formulated based on the fitting results of the
two linear slopes. 
One additionally has the opportunity to exclude dip noise on each slope, 
i.e., parts of the linear function that are in close approximation to the dip and
hence, likely to become noisy during experiments with high temperature changes. 
This is done similarly as step 3 by excluding those parts for which $|I_{\rm Fit}(\omega)| > |I_{\rm dip}^{\rm ref}(\omega_{min1,2})|$ with $|I_{\rm dip}^{\rm ref} (\omega_{\rm min1,2})| = \mu_{1,2} |I (\omega_{\rm min1,2}) - Y_0|$ and $0 \leq \mu_{1,2} \leq 1$.

\item 
To match the requirement for Eq.~\ref{eq:4pnt}, $I(\omega_-) = I(\omega _+)$ pairs are allocated to one another and the two frequencies $\omega_-$ and $\omega_+$ of the pair, that exhibit the highest collective $\delta \omega$ within the extent of the slopes, are chosen as $\omega_{-}$ and $\omega_{+}$.

\item 
The four frequency points, two on each slope, are uniformly distributed, i.e., equally
distanced with $\delta \omega$ from $\omega_{-}$ and $\omega_{+}$, as depicted in Fig.~\ref{fig-fpointselction}. 

\end{enumerate}

\section{Derivation of the fluorescence intensity at four frequency points}
\label{A:B}
The ODMR intensities $I(\omega)$ at the four frequency points can be written as
\begin{equation}
\begin{split}
    I_1 & = I(\omega_-)+\gamma_1 \left[-{\delta \omega}+{\delta \beta}+\delta T  \left(\frac{dD}{dT}\right) \right], \\
    I_2 & = I(\omega_-)+\gamma_1 \left[+{\delta \omega}+{\delta \beta}+\delta T \left(\frac{dD}{dT}\right) \right], \\
    I_3 & = I(\omega_+)+\gamma_2 \left[-{\delta \omega}-{\delta \beta}+\delta T \left(\frac{dD}{dT}\right)\right], \\
    I_4 & = I(\omega_+)+\gamma_2 \left[+{\delta \omega}-{\delta \beta}+\delta T \left(\frac{dD}{dT}\right) \right], \\
\end{split}
\end{equation}
where $\gamma_1$ and $\gamma_2$ depict the slopes of the two linear domains and $dD/dT=\alpha$ describes the temperature dependence of $D$.
${\delta \beta}$ is an unknown static magnetic field~\cite{kucsko2013nanometre} but is assumed to be zero in this investigation.
We assumed that $|\gamma_1|$ and $|\gamma_2|$ are equal; however, they exhibit slight differences ($\sim 50 \%)$, as shown in Table~\ref{tbl3}.

\begin{table}[th!]
\small
  \caption{Variation of $\gamma_1$ and $\gamma_2$ in the 4-point selection process.}
  \label{tbl3}
 \begin{tabular}{cccc}
   \hline
    Sample ND & $\gamma_1$ [MHz$^{-1}$] & $\gamma_2$ [MHz$^{-1}$] & Difference [\%] \\
    \hline
    \textbf{1}  &   -4.821    &   4.594     &   4.7 \\ 
    \textbf{2}  &   -8.807    &   9.112     &   3.3 \\ 
    \textbf{3}  &   -6.152    &   5.825     &   5.6 \\ 
    \textbf{4}  &   -9.326    &   8.825     &   5.4 \\ 
    \textbf{5}  &   -4.440    &   4.194     &   5.9 \\ 
    \hline
 \end{tabular}
\end{table}

\newpage
\section{Calibration of $T_{S}$ as a function of $T_{\rm obj}$}
\begin{figure}[h!]
 \centering
 \includegraphics[scale=0.85]{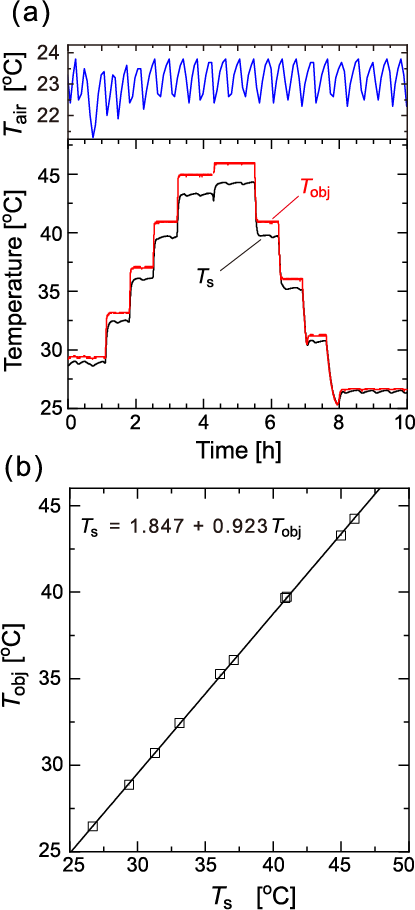}
 \caption{
 Calibration of the sample temperature ($T_{\rm S}$) with that of the microscope objective ($T_{\rm obj}$). (a) Temperature profiles of environemental temperature  ($T_{\rm air}$, blue), $T_{\rm obj}$ (red), and $T_{\rm S}$ over 10 h. (b) The obtained relation between $T_{\rm obj}$ and $T_{\rm S}$ with the linear fit. $T_{\rm S} = 1.847 + 0.923 T_{\rm obj}$ is obtained.
 }
 \label{figS-Tscalib}
\end{figure}

Figure~\ref{figS-Tscalib} shows the calibration data of $T_{\rm S}$ relative to $T_{\rm obj}$.

\clearpage
\onecolumngrid
\section{Error propagation of the second-order-polynomial fit to the temperature estimation}
\begin{figure*}[h!]
 \centering
 \includegraphics[scale=1.0]{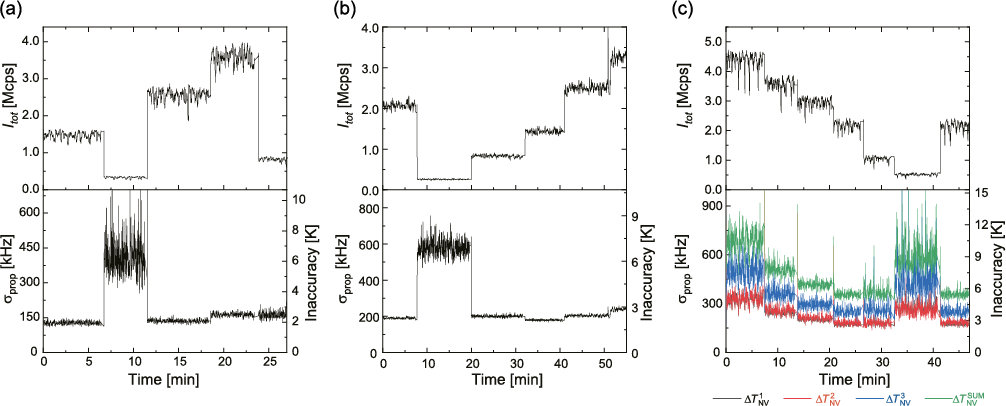}
 \caption{Error propagation to the temperature estimation of $\Delta T_{\rm NV}$. $I_{\rm tot}$ (top) and propagated errors, $\sigma_{\rm prop}$ (bottom) for the (a) 4-point, (b) 3-point, and (c) 6-point measurements.
 }
 \label{fig-errorprop}
\end{figure*}

The errors of the fitting parameters summarized in Table~\ref{tbl0} propagate to the temperature estimation using Eqs.~\ref{eq:4pnt}--\ref{eq:6pnt-2}, which can be explicitly written as Eqs.~\ref{eq:prop4pnt}--\ref{eq:prop6pnt}. 
Figure~\ref{fig-errorprop} shows the propagated errors associated with the time trace in Fig.~\ref{fig5-8}(a). 

\begin{align}
\sigma_{\rm prop}^{\rm 4pnt} =& \frac{2 \delta \omega}{(I_1^C - I_2^C - (I_3^C - I_4^C))^2} \times \nonumber \\
&\sqrt{(I_4^C - I_2^C)^2 \left[ \delta a_0^2 + (I_3^{\rm NC})^2 (\delta a_1^2 + (I_3^{\rm NC})^2 \delta a_2^2)\right] + (I_1^C - I_3^C)^2 \left[\delta b_0^2 + (I_4^{NC})^2 (\delta b_1^2 + (I_4^{NC})^2 \delta b_2^2)) \right]}. \label{eq:prop4pnt} \\
 \nonumber  \\
\sigma_{\rm prop}^{\rm 3pnt} =& -\Gamma \frac{1+\rho^2}{2\rho}
\sqrt{\frac{4(I_0 - I_-)^2 (\delta c_0^2 + (I_+^{\rm NC})^2(\delta c_1^2 + (I_+^{\rm NC})^2\delta c_2^2)}{(2I_0 - I_- -I_+^{\rm C})^4}}. \label{eq:prop3pnt} \\
 \nonumber \\
\sigma_{\rm prop}^{\rm 6pnt} =& \sqrt{\left( \sigma_{\rm prop}^{\rm 4pnt} (I_1, I_2, I_3, I_4)\right)^2 + \left( \sigma_{\rm prop}^{\rm 4pnt} (I_1, I_-, I_+, I_4)\right)^2 + \left( \sigma_{\rm prop}^{\rm 4pnt} (I_-, I_2, I_3, I_+)\right)^2}. \label{eq:prop6pnt}
\end{align}

\clearpage
\twocolumngrid

\section{Three kinds of stepwise temperature profiles in the 6-point measurement}
\begin{figure}[th!]
 \centering
 \includegraphics[scale=0.9]{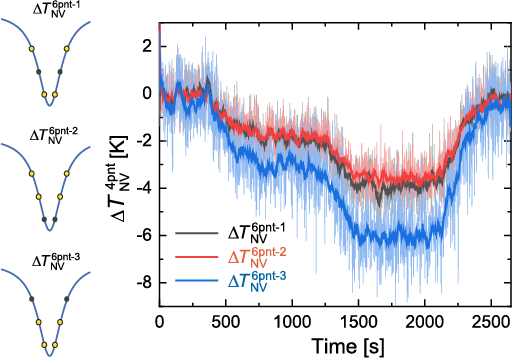}
 \caption{A schematic illustration of the choses points for the 6-point method and the resultant time profiles of $\Delta_{\rm NV}^{1--3}$ during the stepwise temperature change shown in Fig.~\ref{fig-comp}.}
 \label{fig-6pnt-temp}
\end{figure}

Figure~\ref{fig-6pnt-temp} shows the time profiles of each 4-point measurement conducted in the 6-point measurements. 
$\Delta_{\rm NV}^{\rm 6pnt-3}$ that probes the bottom region overestimates the temperature change by a factor of 1.7 compared to the other estimates, which indicates that the bottom region is more shifted than the base region of ODMR dip.

\section{Effect of the spectral shape to the temperature estimation in the 3-point method}
\begin{figure}[th!]
 \centering
 \includegraphics[scale=1]{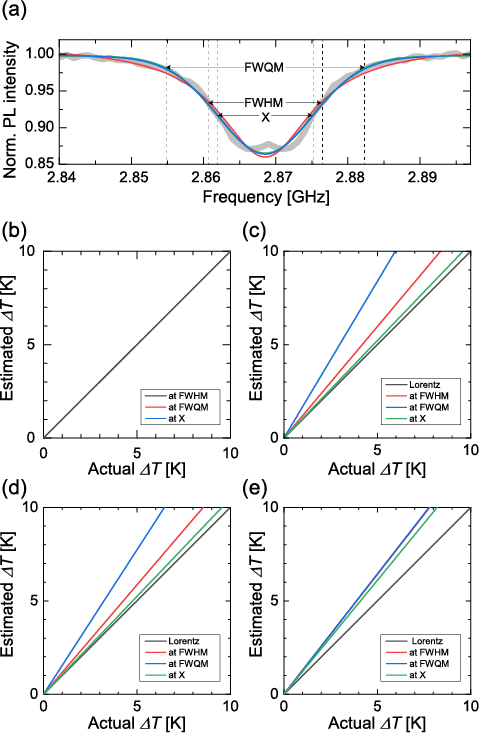}
 \caption{(a) The ODMR spectrum used in Fig.~\ref{fig-comp} is fitted by a single Lorentzian (red), Gaussian (blue), and pseudo-Voght (gree) function. The theoretical analysis on the relationship between the estimated $\Delta T$ and the actual $\Delta T$, for the cases when the ODMR spectrum is regarded as (b) Lorentzian, (c) Gaussian, (d) pseudo-Voght, and (e) interpolating function, respectively.}
 \label{fig-3pnt-theory}
\end{figure}

In the 3-point method, the real spectral shape of the ODMR does not exactly match the single Lorentzian profile, which causes deviations in the estimation from the real temperature change. 
Figure~\ref{fig-3pnt-theory} shows how the spectral shape affects the temperature estimates in the 3-point method for the ND used in Fig.~\ref{fig-comp}.
We first formulate spectral functions by fitting the experimentally measured ODMR spectrum (Fig.~\ref{fig-3pnt-theory} (a)) with single Lorentzian, Gaussian, and pseudo-Voigt function.
In addition, we formulate an interpolated function of the experimental spectrum.
Second, we shift these four spectral functions by assuming the temperature change and numerically estimate $\Delta T$ by the 3-point method (Eq.~\ref{eq:3pnt}). 
When the spectral shape is a perfect single Lorentzian, the 3-point method correctly estimates the temperature change as shown in Fig.~\ref{fig-3pnt-theory} (b).
However, the 3-point method exhibits a significant deviation from the Lorentzian-based estimation when the spectral shape is Gaussian or pseudo-Voigt (Figs.~\ref{fig-3pnt-theory} (c) and (d)). 
The deviation increases as the frequencies of $I_{-}$ and $I_{+}$ increase.
In case of the real ODMR spectrum (interpolated function), the deviation is significant. 

\begin{figure*}[th!]
 \centering
 \includegraphics[scale=0.95]{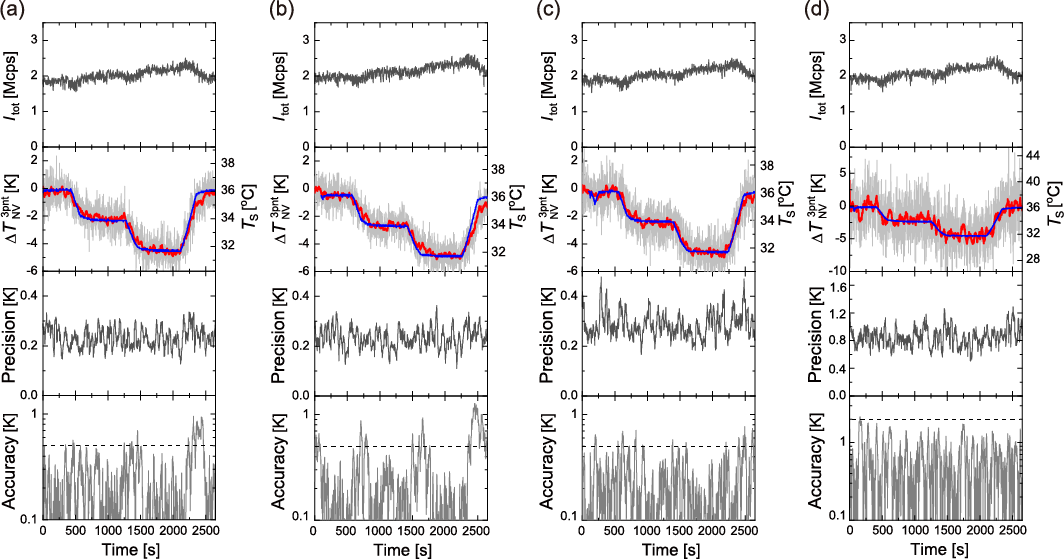}
 \caption{Time profiles of $\Delta T_{\rm NV}$ for the 3-point methods by different experimental parameters. The time profile in (a) the normal configuration (the same as in Fig.~\ref{fig-comp} (b)) and (b) when the microwave at 2.65 GHz is switched off during the measurement of $I_0$. (c) The time profiles when the frequencies of $I_{-}$ and $I_{+}$ are set to FWHM and (d) FWQM (full-width at quarter maximum) (d).} 
 \label{fig-3pnt-comp-tchange}
\end{figure*}

\begin{figure}[th!]
 \centering
 \includegraphics[scale=1]{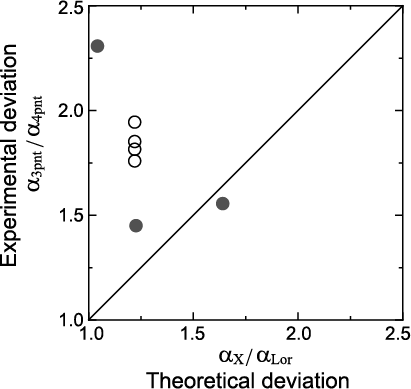}
 \caption{The experimental deviation of $\alpha_{\rm 3pnt} / \alpha_{\rm 4pnt}$ against the theoretical deviation estimated from the spectral shape of the ODMR ($\alpha_{\rm X} / \alpha_{\rm Lor}$). The line with a slope 1 is the guide for the eye . The open circles are of the same ND but measured in different measurement parameters as shown in Fig.~\ref{fig-3pnt-comp-tchange}.}
 \label{fig-devsumm}
\end{figure}
The off-resonant photon count ($I_{0}$) also affects the temperature estimation. 
Figures~\ref{fig-3pnt-comp-tchange} (a) and (b) show the time profiles of the stepwise temperature change measured by the 3-point method with microwave irradiation at 2.65 GHz is ON or OFF during the measurement of $I_{0}$, respectively. 
The irradiation of far off-resonant 2.65 GHz should provide the same photon count as when no microwave is irradiated; however, $I_{0}$ is slightly increased ($\sim 1 \%$) when the 2.65-GHz microwave is irradiated in the present experiment, 
which results in the increase of the temperature dependency of the zero-field splitting ($\alpha$) from -95 to -105 kHz/K.
Such a change of $I_{0}$ by the microwave irradiation is most probably caused by the high-frequency noise exerting to the piezo stage or other electronic devices because we have detected a very small level of noise effect to electronics of piezo stage (for example, the microwave irradiation seems causing the positional shift of laser focal point on the order of 10 nm).
While it could be possible to completely remove such high-frequency noise, the observed sensitivity of the 3-point method to the variation of $I_{0}$ may affect the measurement when working on biological samples because water has absorption still at 2.65 GHz.
The observed sensitivity of the 3-point method fundamentally comes from the fact that $I_0$ term is not completely cancelled in Eq.~\ref{eq:3pnt} in contrast to the formulation of Eq.~\ref{eq:4pnt}.
In addition to the sensitivity to $I_{0}$, the selection of the two frequency points on the slope of the ODMR dip affects the estimation as shown in Figs.~\ref{fig-3pnt-comp-tchange} (c) and (d) where the two frequencies were set to the FWHM and FWQM points. 
In particular, the measurement at FWQM points is very noisy because of the small contrast of ODMR between the off-resonance and measurement points. 

For the four NDs in total, we plot the ratio of $\alpha$ between the 4- and 3-point methods against the theoretical deviation that can be expected from the spectral shape by temperature shifting of the interpolated function as shown in Fig.~\ref{fig-devsumm}.
The experimental deviation is usually larger than the theoretical deviation.
This trend suggests that there should be other causes to the present excessive estimation of $\alpha$ of the 3-point method. 
A more thorough analysis will be necessary in the future development of the 3-point method.

\newpage
\section{Detailed analysis of Allan variance}

\begin{figure}[h!]
 \centering
 \includegraphics[scale=0.9]{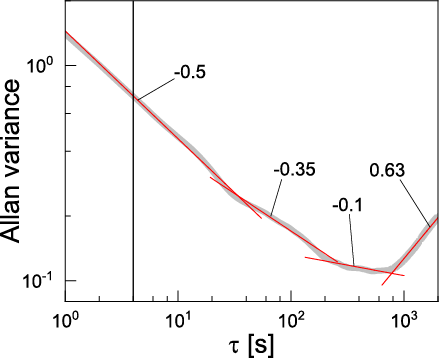}
 \caption{Linear slopes in the Allan variance data of Fig.~\ref{fig-alavar}(b). Slopes of the distinct four regions are indicated. The tracking time of 4 s is indicated as the vertical solid line.  
 }
 \label{figS-alavarslopes}
\end{figure}
\begin{figure}[h!]
 \centering
 \includegraphics[scale=1.0]{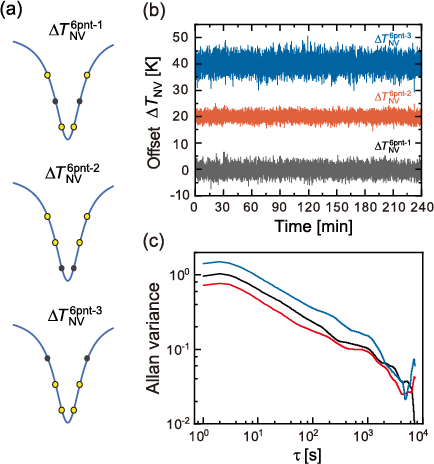}
 \caption{Supplementary data of The Allan variance of the 6-point method. (a) The three types of the 4-point configurations are schematically depicted. (b) The time profiles of $\Delta T_{\rm NV}$ with an offset of 20 K. (c) The Allan variance profile of the respective $\Delta T_{\rm NV}$.
 }
 \label{figS-alavar6pnt}
\end{figure}
Figure~\ref{figS-alavarslopes} shows the slopes of the linear fits to each stage of Allan variance data for the 4-point measurement of Fig.~\ref{fig-alavar}(b).
For a more detailed understanding of the 6-point analysis, we plot time profiles of $\Delta T_{\rm NV}^{\rm 6pnt-1}$, $\Delta T_{\rm NV}^{\rm 6pnt-2}$ and $\Delta T_{\rm NV}^{\rm 6pnt-3}$ for the data shown in Fig.~\ref{figS-alavar6pnt}(c).
The noise level varies for the three types of $\Delta T_{\rm NV}^{\rm 6pnt}$. $\Delta T_{\rm NV}^{\rm 6pnt-3}$ yields the smallest noise and the noise increased in the order of $\Delta T_{\rm NV}^{\rm 6pnt-1}$ and $\Delta T_{\rm NV}^{\rm 6pnt-2}$.
The Allan variances of these three types are shown in Fig.~\ref{figS-alavar6pnt} (f). 
Although the magnitude of $\sigma (\tau)$ is different for these three types, their variance profiles are the same, which indicates that they share the common noise sources.
It should be noted that the present 6-point method provides the same precision compared to the 4-point method because of the estimation function for the frequency shift of ODMR line. 
Rather, the 6-point method is used to understand the experimental hardware necessary for the multipoint measurements.
In principle, the 6-point method provides more information than the 4-point analysis and may be useful to estimate the change of whole spectral shape of ODMR if one uses some designated equations. 

\section{Effect of the tracking period}
\label{sec:ApH}
\begin{figure}[th!]
 \centering
 \includegraphics[scale=0.8]{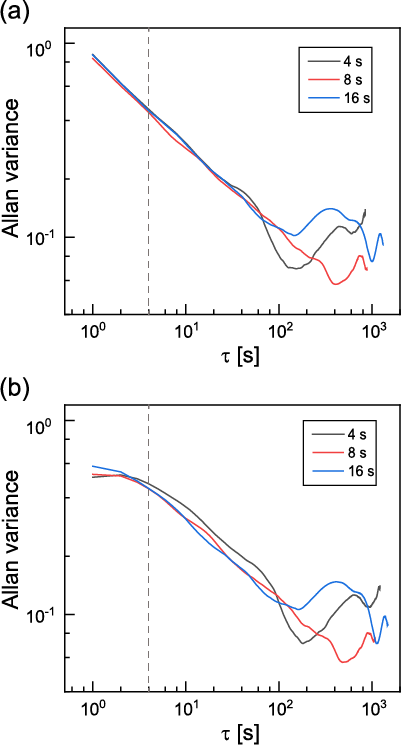}
 \caption{Allan variance profiles in the 4-point method with the tracking period of 4, 8, and 16 s for the two types of measurements in which the tracking times are (a) ignored or (b) included via interpolation (b).
 }
 \label{figS-tracking-time}
\end{figure}

In the main text, the positional tracking is performed every $t_{\rm track} = 4$ s. 
This tracking time can be varied if the positional drift is sufficiently small during the tracking period. 
Figures~\ref{figS-tracking-time} (a) and (b) show Allan variance profiles of the thermometry stability data for $t_{\rm track} =$ 4, 8, 16 s where the interval time is included by interpolation and ignored to regard it in 1-s constant sampling time, respectively. 
Interestingly, all of them exhibit the flat noise profile until 4 s regardless the tracking period as seen in Fig.~\ref{figS-tracking-time}(a) and, in case of ignoring the tracking time, they show almost the same profile until 80 s as in Fig.~\ref{figS-tracking-time} (b). 
The noise profiles between 4 and 80 s differ when including the tracking period via interpolation as in Fig.~\ref{figS-tracking-time} (a). 
These results indicate that the first plateau of noise profile until 4 s is due to the uneven time spacing of the data in the Allan variance calculation or interpolation effect. 

\section{Noise filtering for correct temperature estimation}
\label{A:F}
\begin{figure}[th!]
 \centering
 \includegraphics[scale=0.8]{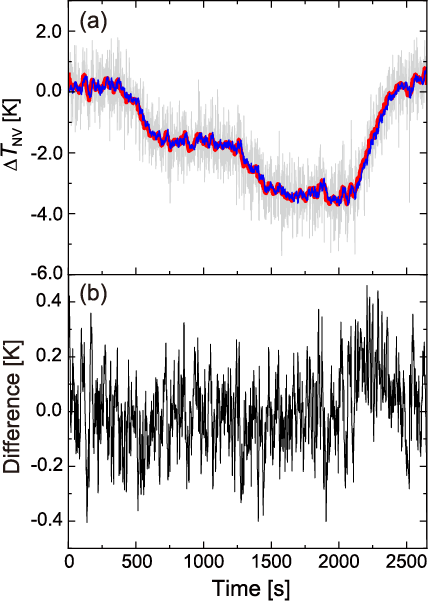}
 \caption{(a) Time profile of $\Delta T_{\rm NV}^{\rm 4pnt}$ (gray) with 20-point moving average (red) and Kalman filter (blue). In Kalman filtering, $\sigma_p^2 =0.01$ and $\sigma_m^2 =1$ were used. (b) The difference between the moving average and Kalman filtering.
 }
 \label{fig-kalman}
\end{figure}

The multipoint ODMR measurements are essentially an estimation of the temperature-dependent frequency shift of ODMR, which motivates us to implement other noise filters and data estimation techniques for such transient dataset.
A suitable choice is the Kalman filter, which has been widely used in systems and control engineering~\cite{4101411kalman-wiener}. We employ the following (one-dimensional) Kalman filter: 
\begin{equation}
\begin{split}
   \hat{x}_{k+1} &= \hat{x}_k^{\prime} + G(z_k - \hat{x}_k^{\prime}) \\ 
   p_{k+1} &= p_k +\sigma_p^2, \\
   G &= \frac{p_k^{\prime}}{p_k^{\prime} + \sigma_m^2}, \\
   p_k &= p_k^{\prime} (1-G),   \\
\end{split}
\end{equation}
where $\hat{x}_{k}$, $\hat{x}_k^{\prime}$, $G$, $z_k$, $p_k$, $p_k^{\prime}$, $\sigma_p^2$, and $\sigma_m^2$ are an estimate of the state $x$ at time $k$, prior estimate of $\hat{x}_{k}$, Kalman gain, actual value of $x$ at time $k$, error covariance at time $k$, prior estimate of $p_k$, and the noise covariance of the prediction and measurement, respectively.
The measurement and prediction errors are considered to be Gaussian.
Figure~\ref{fig-kalman} compares the time profiles of $\Delta T_{\rm NV}$ with the 20-point moving average and the Kalman filter with $\sigma_p^2 =0.01$ and $\sigma_m^2 =1$. 
The two types of filters effectively extract the dynamics of temperature in the ND quantum thermometry and match each other. While the 20-point moving average filter is sufficient to extract the temperature dynamics as long as working on this kind of step temperature change, the successful implementation of the Kalman filter should be useful for more realistic transient temperature dynamics that we cannot predict easily.
Furthermore, a recent demonstration of active feedback quantum thermometry by coupling with a magnetic field may harness the rapid estimation protocol of the Kalman filter.  

\clearpage
\bibliography{achemso-demo}
\bibliographystyle{apsrev4-1}

\end{document}